\documentclass[12pt]{article}
\usepackage{amsmath, amssymb}
\usepackage{graphicx}
\usepackage{natbib}
\usepackage{url} 
\usepackage{algorithm}
\usepackage{algpseudocode}
\usepackage{microtype}
\usepackage{times}
\usepackage{graphicx}
\usepackage{caption}
\usepackage{subcaption}
\usepackage{booktabs}
\usepackage{multirow}
\usepackage{xcolor}
\usepackage{comment}
\usepackage{caption}
\usepackage{mathtools}
\usepackage{xr}
\makeatletter
\newcommand*{\addFileDependency}[1]{
  \typeout{(#1)}
  \@addtofilelist{#1}
  \IfFileExists{#1}{}{\typeout{No file #1.}}
}
\makeatother

\newcommand{\blind}{0}

\newtheorem{theorem}{Theorem}
\newtheorem{assump}{}
\newtheorem{remark}{Remark}

\DeclarePairedDelimiter\floor{\lfloor}{\rfloor}

\begin{document}

\def\spacingset#1{\renewcommand{\baselinestretch}%
{#1}\small\normalsize} \spacingset{1}
\newcommand{\tblue}{\textcolor{blue}}
\newcommand{\linbo}{\textcolor{orange}}


\if0\blind
{
  \title{\bf Simultaneous Estimation of Multiple Treatment Effects from Observational Studies}
  \author{Xiaochuan Shi, Dehan Kong, and Linbo Wang\thanks{
    The authors gratefully acknowledge funding from the Natural Sciences and Engineering Research Council of Canada and Canadian Statistical Sciences Institute for supporting this research.}\hspace{.2cm}\\
    Department of Statistical Sciences, University of Toronto}
    \date{}
  \maketitle
} \fi

\if1\blind
{
  \bigskip
  \bigskip
  \bigskip
  \begin{center}
    {\LARGE\bf Simultaneous Estimation of Multiple Treatment Effects from Observational Studies}
\end{center}
  \medskip
} \fi

\bigskip
\begin{abstract}

Unmeasured confounding presents a significant challenge in causal inference from observational studies. Classical approaches often rely on collecting proxy variables, such as instrumental variables. However, in applications where the effects of multiple treatments are of simultaneous interest, finding a sufficient number of proxy variables for consistent estimation of treatment effects can be challenging. Various methods in the literature exploit the structure of multiple treatments to address unmeasured confounding. In this paper, we introduce a novel approach to causal inference with multiple treatments, assuming sparsity in the causal effects. Our procedure autonomously selects treatments with non-zero causal effects, thereby providing a sparse causal estimation. Comprehensive evaluations using both simulated and Genome-Wide Association Study (GWAS) datasets demonstrate the effectiveness and robustness of our method compared to alternative approaches.




\end{abstract}

\noindent%
{\it Keywords:} Causal inference; Identification; Sparsity assumption; Unmeasured confounding.
\vfill

\newpage
\spacingset{1.75} 
\section{Introduction}
\label{sec:intro}

Causal inference plays a crucial role in addressing research problems across fields such as criminal justice, higher education, and the labor market. 
While randomized experiments are the gold standard for establishing causation in statistical research, ethical constraints or impracticality often preclude their implementation. Consequently, researchers frequently resort to observational data to address causal questions. However, a persistent challenge in observational studies is the presence of unmeasured confounding. In scenarios involving a single treatment, causal identification under unmeasured confounding often relies on  external variables, such as instrumental variables \citep{IV1, wang2018bounded}, and negative controls \citep{proxyvariables}. 

In numerous real-world applications such as Genome-Wide Association Studies (GWAS) and computational biology, multiple treatments instead of a single treatment are of interest. A simple generalization from a single treatment to multiple treatments is to consider the treatments as a single vector-valued treatment. For example, \cite{damourDiscuss} and \cite{ImaiDiscuss} suggest extending negative controls and instrumental variable approaches to accommodate multi-treatment scenarios. However, extending existing approaches to the multi-treatment setting is complicated by the fact that the instrumental variable approach often requires as many IVs as there are treatments, and the negative controls approach necessitates at least two confounder proxies for each treatment. Attaining this can be challenging in many observational studies.

The literature has seen a growing interest in methods specifically tailored for multi-treatment settings without requiring additional auxiliary variables. Most notably, \cite{deconfounder} introduces the so-called deconfounder method, which employs factor models to estimate a substitute for unmeasured confounders. Nonetheless, as first pointed out by \cite{damourDiscuss}, the deconfounder approach does not guarantee the identification of causal effects, meaning that the treatment effects cannot be uniquely determined from the observed data under their assumptions. \cite{kong} show that under a linear Gaussian model for the treatments and a binary choice model for the outcome with a non-probit link, the causal effects are identifiable. Under the linear structural equation model for the outcome, \cite{Miao} identify the causal effects by assuming sparsity for the causal effects of multiple treatments. Crucially, they identify and estimate the causal effects without estimating the unmeasured confounder. More recently, \cite{cevid2020spectral} and \cite{sparRegress} proposed consistent estimation methods for the causal effects in high-dimensional multi-treatment settings under a stricter sparsity assumption than that in \cite{Miao}.


In this paper, we present a novel approach for  simultaneous estimation of multiple causal effects under a sparsity assumption of these effects. Sparsity is frequently assumed when the number of parameters to be estimated exceeds the sample size. Moreover, the sparsity assumption contributes to result interpretability \citep{interpretableSpar}. \cite{Miao} established identifiability of the causal effects under the assumption that no more than $(p-q)/2$ causal effect entries are non-zero. Here, $p$ is the number of treatments, and $q$ denotes the dimension of unmeasured confounders. However, the least median squares estimator in \cite{Miao} is consistent only when at most $\floor*{p/2} - q + 1$ entries of causal effects are nonzero, where $\floor*{x}$ is the largest integer less than or equal to $x$. This is a stronger requirement compared to their identification condition. Our estimation approach contributes to the current literature on estimating causal effects for multiple treatments in three aspects. First, it bridges the gap in the sparsity assumptions between identification and estimation in \cite{Miao}, requiring the same level of sparsity for both aims. We achieve this through a novel optimization-based identification approach, which also forms the basis of our estimation method. Second, our estimation procedure automatically selects the treatments with non-zero causal effects on the outcome, resulting in a sparse estimator. Finally, our method is capable of addressing both low and high dimensional scenarios, where the number of treatments may or may not exceed the number of samples.

The remainder of this paper is structured as follows. In Section \ref{sec:pre}, we introduce basic notations and formulate the estimation problem. In Section \ref{sec:iden}, we outline the assumptions under which the causal effects can be identified. The estimation procedure is detailed in Section \ref{sec:esti}. In Sections \ref{sec:simu} and \ref{sec:gwas}, we evaluate our method and compare it with alternative approaches using simulations and synthetic GWAS datasets. We conclude with a brief discussion in Section \ref{sec:discuss}.

\section{Method}
\label{sec:method}


\subsection{Preliminary}
\label{sec:pre}

Let $X$ represent the $p$-dimensional treatments, and $Y$ denote the real-valued outcome. Let $U$ denote the $q$-dimensional unobserved confounders that may influence both the treatments and the outcome, and $W$ denote the $r$-dimensional measured confounders.
We consider the following linear structural equation models \citep{pearlCausal}:
\begin{equation}
\label{equ:model}
    \begin{aligned}
    &X = \alpha U +\eta W + \epsilon_x ,\\
    &Y = \beta^T X + \delta^T U +\lambda^T W + \epsilon_y,
    \end{aligned}
\end{equation}
where $\alpha \in \mathcal{R}^{p\times q}$, $\beta \in \mathcal{R}^p$, $\delta \in \mathcal{R}^q$, $\eta \in \mathcal{R}^{p\times r}$, and $\lambda \in \mathcal{R}^r$ are coefficients. The random errors $\epsilon_x \in \mathcal{R}^p$, $\epsilon_y \in \mathcal{R}$, and confounders $U$ are independent of each other.  Denote $\Sigma_{\epsilon_x}$ as the covariance matrix of $\epsilon_x$, and $\Sigma_U$ as the covariance matrix of $U$. Without loss of generality, we assume that $E(\epsilon_x) = 0$, $E(U) = 0$, $\Sigma_U = I_q$. For simplicity, we omit the measured confounders $W$ from the model in the rest of the paper, as one can adjust for them by regressing both the treatments and the outcome on $W$ and then using the regression residuals as $X$ and $Y$. In Section E of the supplementary material, we present simulation results for both low and high dimensional settings, incorporating the measured confounder $W$. The results show that the performance of our method remains consistent, both with and without the inclusion of $W$. Following the potential outcome framework, we assume that $Y(x)$, the potential outcome under an  intervention that sets $X=x$, is well-defined. The causal effect is defined as the difference between expected potential outcomes given two treatment values. Throughout the paper, we make three standard identifying assumptions: (i) \textit{Consistency}: \( Y = Y(x) \) when \( X = x \). (ii) \textit{Latent ignorability}: For any \( x \), \( X \perp Y(x) \mid U \). (iii) \textit{Positivity}: \( 0 < P(X = x \mid U = u) < 1 \) for all \( (x, u) \) almost surely. Under model (\ref{equ:model}), we have $E(Y(x)) = E(E(Y|X=x,U)) = E(\beta^Tx + \delta^T U + \epsilon_y) = \beta^T x$. Therefore, the causal effect of the treatments $X$ on the outcome $Y$ can be characterized by the parameter $\beta$.

By plugging the first equation into the second in linear model (\ref{equ:model}), we get the marginal {association between} $X$ {and} $Y$:
\begin{equation}
    \begin{aligned}
        &Y = \xi^T X + \epsilon_y^{'},\\
        &\xi = \beta + \gamma\delta, \;\; \gamma = \Sigma_X^{-1}\alpha, \;\; Cov(X, \epsilon_y^{'}) = 0,
    \end{aligned}
    \label{equ:xi}
\end{equation}
where $\Sigma_X$ is the covariance matrix of $X$, and $\epsilon_y^{'} = \delta^T(I - \alpha^T\Sigma_X^{-1}\alpha)U - \delta^T\alpha^T\Sigma_X^{-1}\epsilon_x + \epsilon_y$. 

While linear regression of $Y$ on $X$ provides a consistent estimate of the marginal association, denoted as $\xi$, our primary interest lies in estimating the causal parameter $\beta$. When $\alpha\delta \neq 0$, the association $\xi$ does not equal the causal effect $\beta$ due to confounding. When $U$ is unobservable, the parameters $\alpha$ and $\delta$ cannot be uniquely identified from the joint distribution of $(X,Y)$. We now introduce additional assumptions crucial for establishing the identifiability of the parameter $\beta$.

\subsection{Identification}
\label{sec:iden}

In this subsection, we discuss  
identifiability of the causal effect parameter $\beta$. We first revisit identification results on the loading matrix $\alpha$ in the factor model $X = \alpha U + \epsilon_x$. This problem has been extensively studied within the factor analysis literature \citep{factana,PCAhidimfact, correlateHighd}. When the dimension of the latent factors $q$ is known, it is established that $\alpha$ can be identified up to a rotation under Assumptions A1 and A2 outlined later in this subsection \citep[e.g.,][]{factana}. Given that $\gamma = \Sigma_X^{-1}\alpha$, it follows that $ \gamma $ is also identifiable up to rotation. The identification of $ \xi$ is accomplished through linear regression:
\begin{equation}
\label{equ:lr}
    \xi = \underset{\tilde\xi}{\arg\min} \; E(Y - \tilde\xi^T X)^2.
\end{equation}
The following assumptions are useful for identifying the causal parameter $\beta$.




\begin{assump}
    $\Sigma_{\epsilon_x}$ is diagonal.
\end{assump}

\begin{assump}
    After deleting any row, there remain two disjoint submatrices of $\alpha\in \mathcal{R}^{p\times q}$ of rank $q$.
\end{assump}

\begin{assump}
    Any submatrix of $\gamma \in \mathcal{R}^{p\times q}$ consisting of $q$ rows and $q$ columns has full rank $q$.
\end{assump}

\begin{assump}
    (Sparsity) At most $(p-q)/2$ entries of $\beta$ are nonzero.
\end{assump}

Assumption $A1$ requires the independence of noises across distinct treatment variables. Assumption $A2$ implies that $p \geq 2q + 1$, indicating that the dimension of unmeasured confounders is relatively small compared to the dimension of the treatment $p$. Both Assumptions $A1$ and $A2$ are classical conditions for identification of the loading matrix $\alpha$ in the factor analysis literature \citep{factana}. Under model (\ref{equ:model}) and Assumptions $A1$ \& $A2$, the factor loading $\alpha$ is identifiable up to a rotation, as stated in Theorem 5.1 of \cite{factana}. In cases where treatments remain correlated even after conditioning on the latent factors, Assumption $A1$ can be relaxed if the dimension of treatments $p$ is large. Detailed discussions on the relaxation of Assumption $A1$ can be found in \cite{PCAhidimfact} and \cite{correlateHighd}. Assumption $A3$ is a testable condition with an estimator of $\gamma$. 

Our identification approach is built on the relation
\begin{equation}
    \label{eqn:key}
       \beta = \xi - \gamma\delta,
\end{equation}
where $\xi$ is identifiable, and $\gamma$ is identifiable up to a rotation. Assumption $A4$ implies that $\beta$ is ``sufficiently'' sparse: $\Vert\beta\Vert_0 \leq (p-q)/2$. On the other hand, Assumption $A3$ guarantees that there cannot be more than one solution to \eqref{eqn:key} that is sufficiently sparse. Suppose instead that there exists another $\beta = \xi - \gamma\delta$ that is at least as sparse as $\beta_0$: $\Vert\beta\Vert_0 \leq \Vert\beta_0\Vert_0 \leq (p-q)/2$. By the pigeonhole principle, there are at least \(p - (p-q)/2 - (p-q)/2 = q\) indices for which $\beta - \beta_0 = \gamma(\delta_0 - \delta)$ equals zero. Under Assumption $A3$, this can only occur if $\beta - \beta_0$ is zero.

Following this, in Theorem \ref{them:1} below, we show that under these assumptions, the identification problem of $\beta$ can be formulated as an optimization problem, where the true causal effect $\beta_0 = \xi - \gamma\delta_0$ is the sparsest among all possible $\xi - \gamma\delta$ for any $\delta \in \mathbb{R}^q$.

\begin{theorem}
\label{them:1}
    Under the linear structural model (\ref{equ:model}), and Assumptions $A1$-$A4$, $\gamma$ is identifiable up to a rotation, and $\beta$ is identifiable through the following equations:

    \begin{equation}
    \label{equ:thm1}
        \begin{aligned}
         &\tilde\delta = \arg \min_{\delta} ||\xi -\tilde\gamma  \delta||_0,\\
         &\beta = \xi - \tilde\gamma \tilde\delta,
        \end{aligned}
    \end{equation}where $\tilde\gamma = \gamma R$ for an arbitrary rotation matrix $R$, and $||\cdot||_0$ denotes the $\ell_0$ norm.
\end{theorem}

We defer the proof of Theorem 1 to Section A of the supplementary material. The assumptions in our theorem align with those in the identification result of Theorem 3 in \cite{Miao}. However, our identification strategy, as described in (\ref{equ:thm1}), through optimization, is novel and paves the way for the development of the estimation procedure detailed in Section \ref{sec:esti}.

\begin{remark}
    Assumption $A3$ requires that all treatments are confounded by $U$, as a row of zeros in $\gamma$ would cause any submatrix containing this row to be rank deficient. However, this condition can be relaxed to Assumption $A3^{'}$: any submatrix of $\gamma$ with $q+r$ rows must have full rank $q$, where $0 \leq r \leq p-q$. Correspondingly, Assumption $A4$ can be modified to Assumption $A4^{'}$: at most $(p-q-r)/2$ entries of $\beta$ are nonzero. Assumptions $A3^{'}$ and $A4^{'}$ allow for unconfounded treatments and by the same logic as in Theorem \ref{them:1}, allow for  identification of the causal effect $\beta$.
\end{remark}

\subsection{Estimation}
\label{sec:esti}

Assume we observe $n$ i.i.d. samples of $(X,Y)$ from model (\ref{equ:model}). In this section, we propose an estimation procedure for the causal effect $\beta$ based on the optimization problem (\ref{equ:thm1}) in Theorem 1. 

In the low-dimensional setting where $n>p$, we estimate $\xi$ using the ordinary least squares (OLS) method by regressing $Y$ on $X$. The factor loading matrix $\alpha$ is estimated using the maximum likelihood method, while $\Sigma_X$ is estimated using the sample covariance matrix.

In the high-dimensional setting where the dimension  of treatments exceeds the sample size (i.e., $p > n$), we employ the de-biased Lasso technique \citep{debiasedLasso} to estimate the regression parameter $\xi$. This choice is motivated by the de-biased Lasso's ability to yield a smaller bias compared to traditional Lasso methods. Additionally, in the high-dimensional context, we utilize the principal component analysis (PCA) method \citep{PCAhidimfact}, which can consistently and efficiently estimate the loading matrix $\alpha$ up to a rotation. For the estimation of the high-dimensional covariance matrix $\Sigma_X$ under the assumed factor model in (\ref{equ:model}), we leverage the principal orthogonal complement thresholding (POET) method introduced in \cite{POET}. This method is specifically designed to provide consistent estimates of high-dimensional covariance matrices. The POET method relies on the assumption that \(\Sigma_{\epsilon_x}\) is approximately sparse, as defined in \cite{approsparse1} and \cite{approsparse2}. Our estimated parameters are denoted as $\widehat{\xi}$, $\widehat{\alpha}$, and $\widehat{\gamma}$, where $\widehat{\gamma} = \widehat{\Sigma}_X^{-1} \widehat{\alpha}$.



By replacing $\xi$ and $\tilde\gamma$ with $\widehat{\xi}$ and $\widehat{\gamma}$ in (\ref{equ:thm1}), we can proceed to solve the following problem:
\begin{equation}
\label{equ:solve}
\begin{aligned}
   &\widehat{\delta} =  \arg \min_{\delta} \sum_{i = 1}^p I(|\widehat{\xi}_i - \widehat{\gamma}_i \delta| > t),\\
   &\widehat{\beta}_i = (\widehat{\xi}_i - \widehat{\gamma}_i \widehat{\delta})I(|\widehat{\xi}_i - \widehat{\gamma}_i \widehat\delta| > t), \; i = 1,\cdots p,
\end{aligned}
\end{equation}
where $\widehat\gamma_i$ denotes the $i_{th}$ row of $\widehat{\gamma}$. When $\xi_i - \tilde\gamma_i \delta$ equals zero, the estimated counterpart $\widehat{\xi}_i - \widehat{\gamma}_i \delta$ may not be precisely zero due to estimation errors. To address this, we introduce a threshold $t$ to accommodate the estimation errors associated with $\widehat{\xi}$ and $\widehat{\gamma}$. The threshold is set as $t = \sqrt{2n^{-1}\log(p)\widehat\sigma^2}$ \citep{threshold} to distinguish non-zero signals from noise. Here, $\widehat\sigma^2$ is set as the mean of the residual variance obtained from the factor analysis of $X$ and the regression of $Y$ on $X$: $\widehat{\sigma}^2 =\left(\text{mean}(\text{diag}(\widehat\Sigma_{\epsilon_x})) + \widehat\sigma_{\epsilon_y^{'}}^2 \right)/2$. In the low-dimensional setting, \(\widehat\sigma_{\epsilon_y^{'}}\) is estimated using the standard deviation of the residuals. In the high-dimensional setting, following the de-biased Lasso estimation procedure in \cite{debiasedLasso}, we calculate the noise variance using the scaled Lasso method  \citep{scaledLasso}.

The minimization problem in (\ref{equ:solve}) involves minimizing a sum of indicator functions, which is non-convex and hence computationally challenging. To make problem (\ref{equ:solve}) tractable, we reformulate it as a mixed-integer programming (MIP) problem \citep{mixedint} as detailed in (\ref{equ:mip}), with the additional assumption that $||\beta||_{\infty} \leq M$ for a known scalar $M$:
\begin{equation}
\label{equ:mip}
    \begin{aligned}
    &(\widehat{z}^{mip}, \widehat{\delta}^{mip}) = \arg \min_{z,\delta}  1^T_p z = \sum_{j=1}^p z_j\\
    &w.r.t. -Mz - t \leq \widehat{\xi} - \widehat{\gamma} \delta \leq Mz +t, z\in \{0,1\}^p.
    \end{aligned}
\end{equation}
We use the GNU Linear Programming Kit (GLPK) to solve the mixed-integer programming problem in (\ref{equ:mip}). An initial estimator for $\beta$ can then be obtained as $\widehat{\beta}^{mip} = (\widehat{\xi} - \widehat{\gamma} \widehat{\delta}^{mip})\widehat{z}^{mip}$. 

The estimator, $\widehat{\delta}^{mip}$, however, is biased due to  the inclusion of the threshold \(t\). To address this,  we use $\widehat\beta^{mip}$ as an initial value to select treatments that have no causal effects on the outcome. Let $\mathcal{I} \subseteq \{1,\cdots p\}$ contain indices corresponding to the smallest $(p + q)/2$ entries of $|\widehat\beta^{mip}|$. We subsequently update the estimates of $\delta$ and $\beta$ as follows: we obtain $\widehat{\delta}$ by regressing $\widehat{\xi}_i$ on $\widehat{\gamma}_i$ for $i \in \mathcal{I}$, leading to $\widehat{\delta} = \left(\widehat{\gamma}_{\mathcal{I}}^T \widehat{\gamma}_{\mathcal{I}}\right)^{-1}\widehat{\gamma}_{\mathcal{I}}^T \widehat{\xi}_{\mathcal{I}}$, where $M_{\mathcal{I}}$ is the submatrix of $M$ containing rows corresponding to the indices in the set $\mathcal{I}$. The final estimator $\widehat{\beta}$ is computed as $\widehat{\beta} = (\widehat{\xi} - \widehat{\gamma} \widehat{\delta})\widehat{z}^{mip}$. This procedure is akin to performing OLS after using Lasso for variable selection; if the variables are correctly selected, the second-step OLS enhances the estimation accuracy. Throughout the remainder of this paper, we refer to our proposed method as the ``Spar'' method, given that the estimator $\widehat\beta$ is sparse and automatically selects the active treatments that have causal effects on the outcome.


The number of unobserved confounders $q$ may be estimated using existing methods in factor analysis \citep{nfact,nfactOnatski,nfactLR}. 
In the synthetic Genome-Wide Association Study (GWAS) data analysis in Section \ref{sec:gwas}, we employ the hypothesis testing method developed in \cite{nfactOnatski} to determine the dimensionality of the latent factors $U$. Our estimation procedure is summarized in Algorithm 1.


\begin{algorithm}[!ht]
  \caption{\quad The Spar method for simultaneous estimation of multiple causal effects}
\label{alg:spar}
\hspace*{0.02in} {\bf Input: }  
 The treatments ${\bf {X}} \in \mathcal{R}^{n\times p}$(centered), outcome ${\bf {Y}} \in \mathcal{R}^{n\times 1}$, and $M$ the upper bound of $||\beta||_{\infty}$
\hspace*{0.02in} 
\begin{algorithmic}[1]  
\State Obtain $\widehat{q}$ \ from $\bf X$ via the Onatski's test. 
 \If{$n>p$} 
 
 obtain $\widehat\alpha$ and $\widehat{\Sigma}_{\epsilon_x}$ via maximum likelihood estimation, and derive $\widehat{\xi}$ and the residual variance $\widehat{\sigma}_{Y\sim X}^2$ by performing a linear regression of $\bf Y$ on $\bf X$. Let $\widehat{\Sigma}_X$ be the sample covariance matrix of $X$.
 \Else 

 estimate $\alpha$  using Principal Component Analysis (PCA), and derive $\widehat{\Sigma}_X$ and $\widehat{\Sigma}_{\epsilon_x}$ employing the POET method \citep{POET}. Utilize the de-biased Lasso method \citep{debiasedLasso} to obtain $\widehat\xi$ and the residual variance $\widehat{\sigma}_{Y\sim X}^2$.

\EndIf

\State Obtain the threshold $t = \sqrt{2n^{-1}log(p)\widehat\sigma^2}$, where $\widehat{\sigma}^2 = \left(mean(diag(\widehat{\Sigma}_{\epsilon_x})) + \widehat{\sigma}_{Y\sim X}^2\right) / 2$, and $\widehat{\gamma} = \widehat{\Sigma}_X^{-1} \widehat{\alpha}$.

\State Solve the mixed integer programming problem (\ref{equ:mip}) using GLPK solver, and obtain $(\widehat{z}^{mip},\widehat\beta^{mip})$.

\State  {Obtain $\widehat{\delta} = \left(\widehat{\gamma}_{\mathcal{I}}^T \widehat{\gamma}_{\mathcal{I}}\right)^{-1}\widehat{\gamma}_{\mathcal{I}}^T \widehat{\xi}_{\mathcal{I}}$, where $\mathcal{I} \subseteq \{1,\cdots p\}$ contains indices corresponding to the smallest $(p + q)/2$ entries of $|\widehat\beta^{mip}|$.} Obtain the Spar estimator $\widehat{\beta} = (\widehat{\xi} - \widehat{\gamma} \widehat{\delta})\widehat{z}^{mip}$.
\end{algorithmic}
\end{algorithm}

\begin{remark}
    As discussed in Section \ref{sec:intro}, the null treatments approach described in \cite{Miao} also estimates the causal effects \( \beta \) under linear models. However, it necessitates a stricter sparsity assumption compared to Assumption $A4$. In the estimation process of the null treatments approach, the parameter \( \delta \) is determined via a least median of squares estimator, which minimizes the median of the squared residuals \( (\xi - \gamma\delta)^2 \). In contrast, our proposed method minimizes the \( \ell_0 \) norm. Furthermore, we extend both our method and the null treatments approach in \cite{Miao} to accommodate high-dimensional settings where \( p > n \).
\end{remark}
\section{Simulations}
\label{sec:simu}

\subsection{Low dimensional case}
\label{sec:lowd}

In this section, we generate three confounders ($U$), 13 treatments ($X$), and an outcome ($Y$) following the linear model (\ref{equ:model}). Let $s$ represent the $\ell_0$ norm of $\beta$. We let $\beta = (1,\cdots 1, 0, \cdots 0)^T$, where the first $s$ entries are set to $1$, and the remaining entries are $0$. Additionally, we set $\delta = (1,1,1)^T$, and the entries of factor loading matrix $\alpha$ are chosen randomly from $Uniform(-1,1)$. The latent confounders $U$ are simulated from $N(0,I_3)$, and the random errors are generated from $\epsilon_x \sim N(0,I_{13})$, $\epsilon_y \sim N(0,1)$. To ensure \( ||\beta||_{\infty} \leq M \) is not overly restrictive, we select a relatively large bound. Experimental results show that estimation performance is stable for \( M \) between 20 and 50. Thus, we set \( M = 30 \) as a practical value   and apply the same value in both the  simulations   and the GWAS experiments (Section \ref{sec:gwas}). Note that Assumptions $A1$-$A3$ are satisfied under the simulation setting described above. For our simulations, we set the sample size at $n=1000$, and vary $s$ from $1$ to $13$.

We denote the null treatment method in \cite{Miao} as Null, and refer to the simple linear regression of $Y$ on $X$ as OLS. For the Null method and our proposed Spar method, we assume that the number of confounders is correctly specified as $q=3$. In Section B of the supplementary material, we explore the performance of the Spar and Null methods under estimated and incorrectly specified $q$ values. Experimental results demonstrate that the performance of the Spar method remains consistent regardless of whether the number of unmeasured confounders is known a priori or estimated through the hypothesis testing method developed in \cite{nfactOnatski}. We investigate the performance of both the Spar and Null methods for different sparsity levels $s$. We run $1000$ Monte-Carlo simulations and report two key metrics: the mean absolute error (MAE, $\sum_{i=1}^p |\widehat{\beta}_i - \beta_i| / n$) and the root-mean-square error (RMSE, $\sqrt{\sum_{i=1}^p (\widehat\beta_i - \beta_i)^2 / n}$).

\begin{figure}[htbp!]
\centering
\begin{subfigure}[b]{.45\textwidth}
    \centering
    \includegraphics[width=\textwidth]{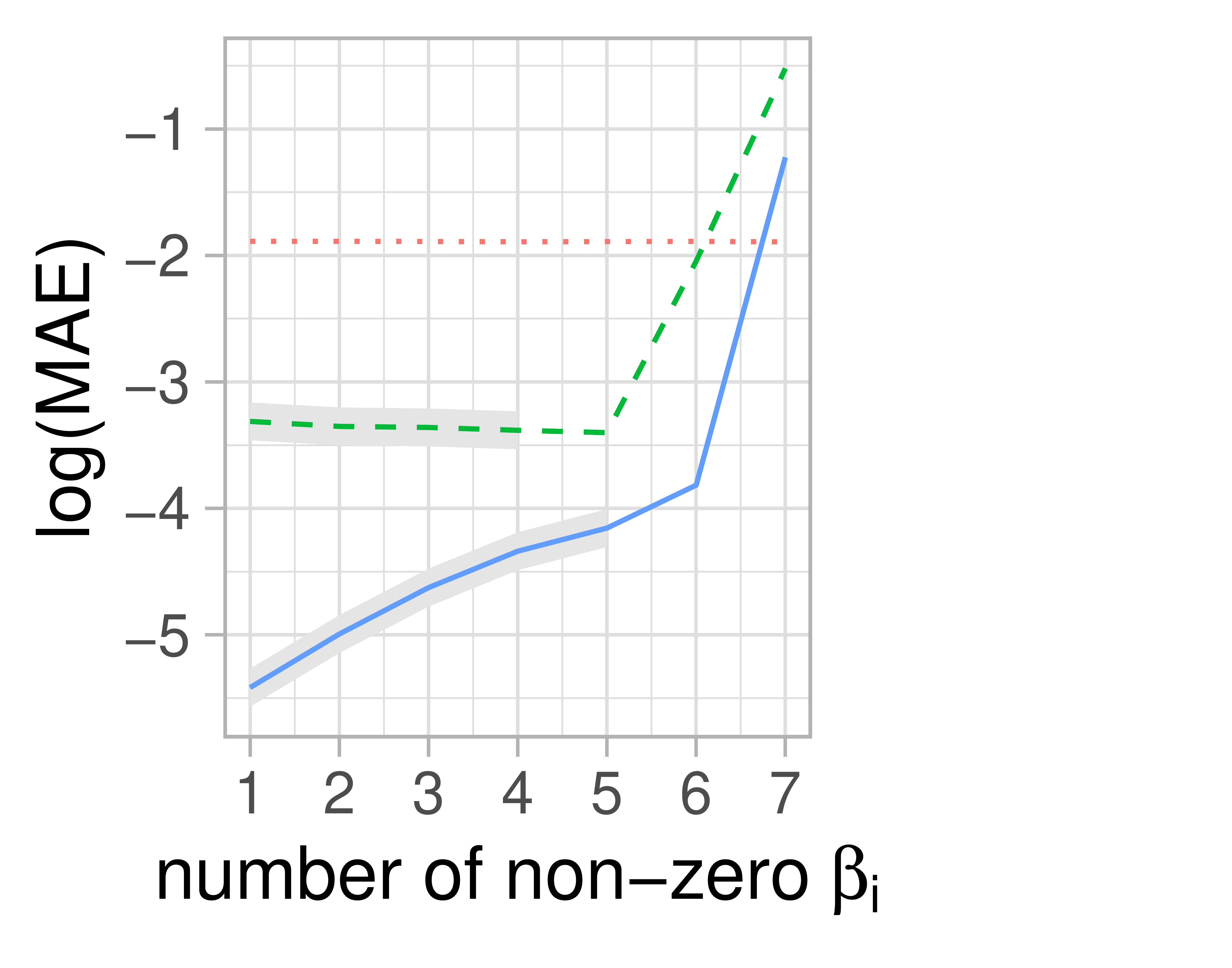}
\end{subfigure}\hspace{-1em}%
\begin{subfigure}[b]{.45\textwidth}
    \centering
    \includegraphics[width=\textwidth]{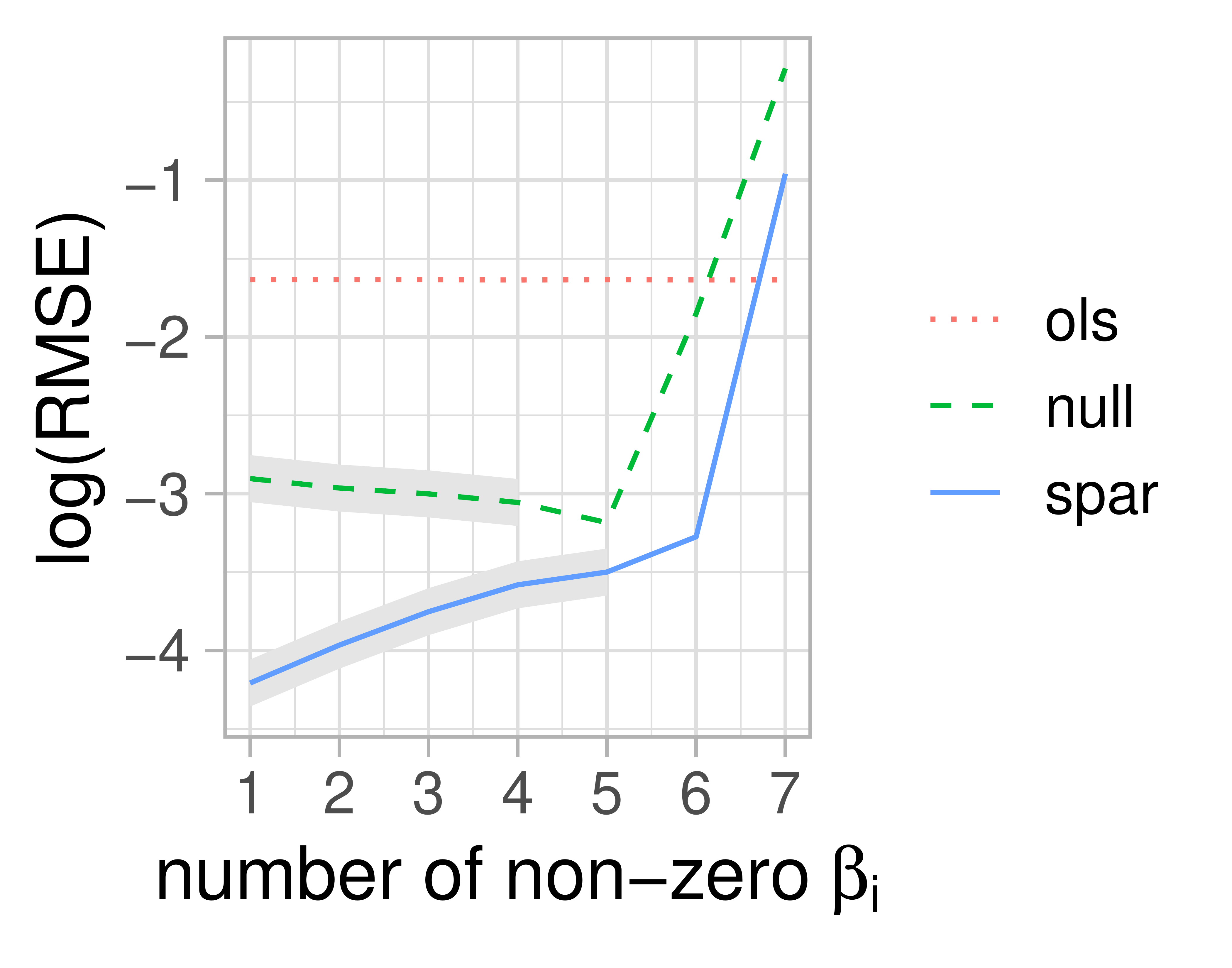}
\end{subfigure}
\caption{The MAE and RMSE of Spar, Null, and OLS methods under different sparsity levels of causal effects $\beta$. The grey area highlights the sparsity levels where the Spar and Null methods are proven to be consistent. The estimation errors increase abruptly beyond the grey area, illustrating the relaxed sparsity constraint of the Spar method in contrast to the Null method.}
\label{fig:errbysparse}
\end{figure}

When $p=13$ and $q=3$, Assumption $A4$ imposes a requirement that $s \leq 5$ in the Spar method, while the Null method is proven to be consistent only when $s \leq 4$. Figure \ref{fig:errbysparse} illustrates the performance of both the Spar and Null methods in terms of estimation errors across various sparsity levels of $\beta$. The figure reveals that the estimation errors of both Spar and Null methods increase as the number of zero entries in $\beta$ decreases. When the number of non-zero entries in $\beta$ exceeds 6, both methods struggle to mitigate the estimation bias induced by unobserved confounders, exhibiting worse performance compared to the simple linear regression. This implies that both methods rely on the sparsity assumption of causal effects. As highlighted by the grey area in Figure \ref{fig:errbysparse}, the Null method maintains relatively low estimation errors until $s > 5$, whereas the Spar method continues to perform well until $s > 6$. This observation is consistent with the theoretical result that the Null method requires a stricter sparsity assumption than the Spar method. We observe that both the Null and Spar methods tolerate slight violations of the sparsity assumptions in our simulation, although such tolerance is not theoretically guaranteed.


\begin{figure}[htbp!]
\vspace{-3mm}
\makebox[10pt]{\raisebox{50pt}{\rotatebox[origin=c]{90}{\small{s=3}}}}%
\begin{subfigure}[b]{.30\textwidth}
    \centering
    \includegraphics[width=\textwidth]{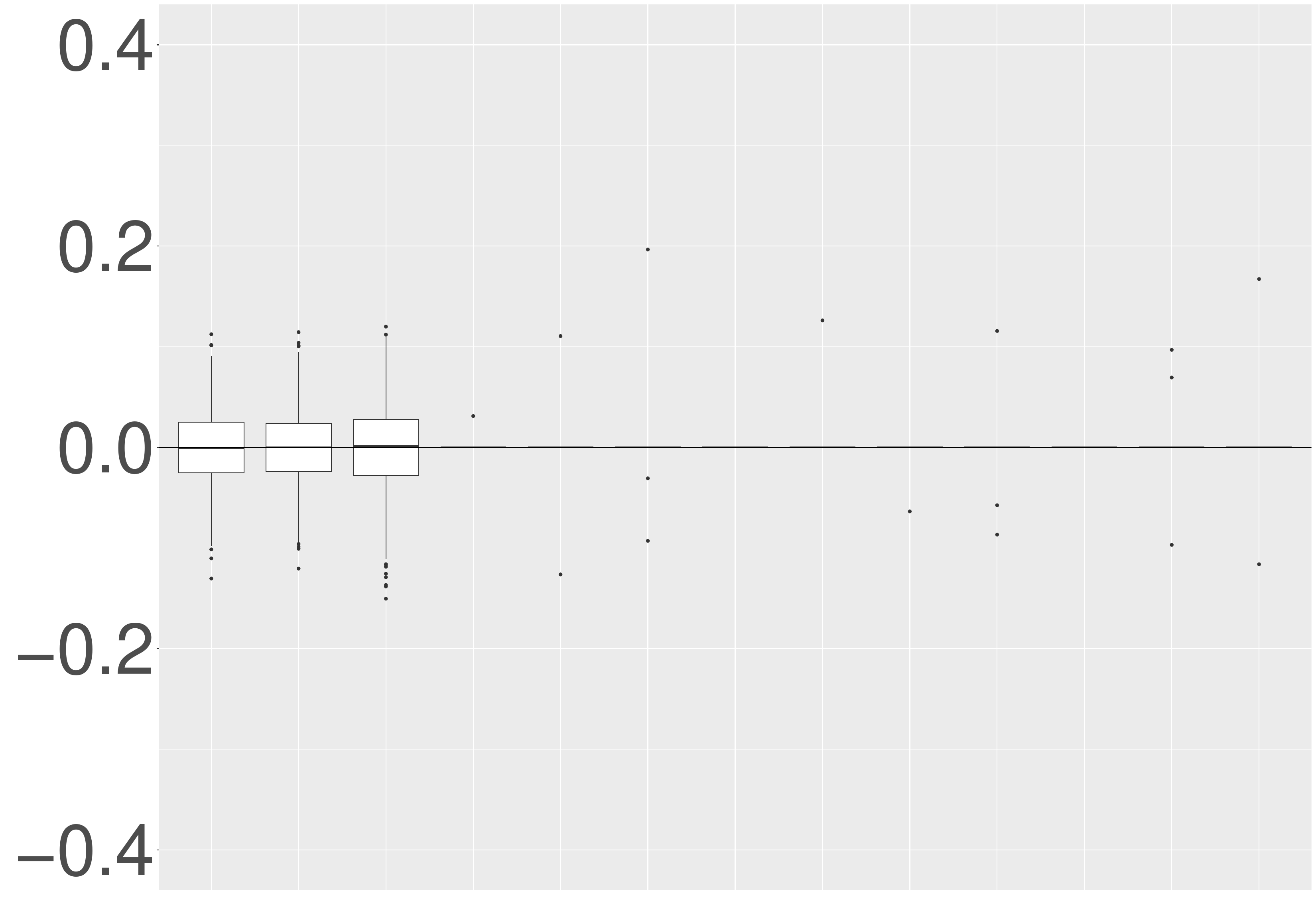}
\end{subfigure}
\begin{subfigure}[b]{.30\textwidth}
    \centering
    \includegraphics[width=\textwidth]{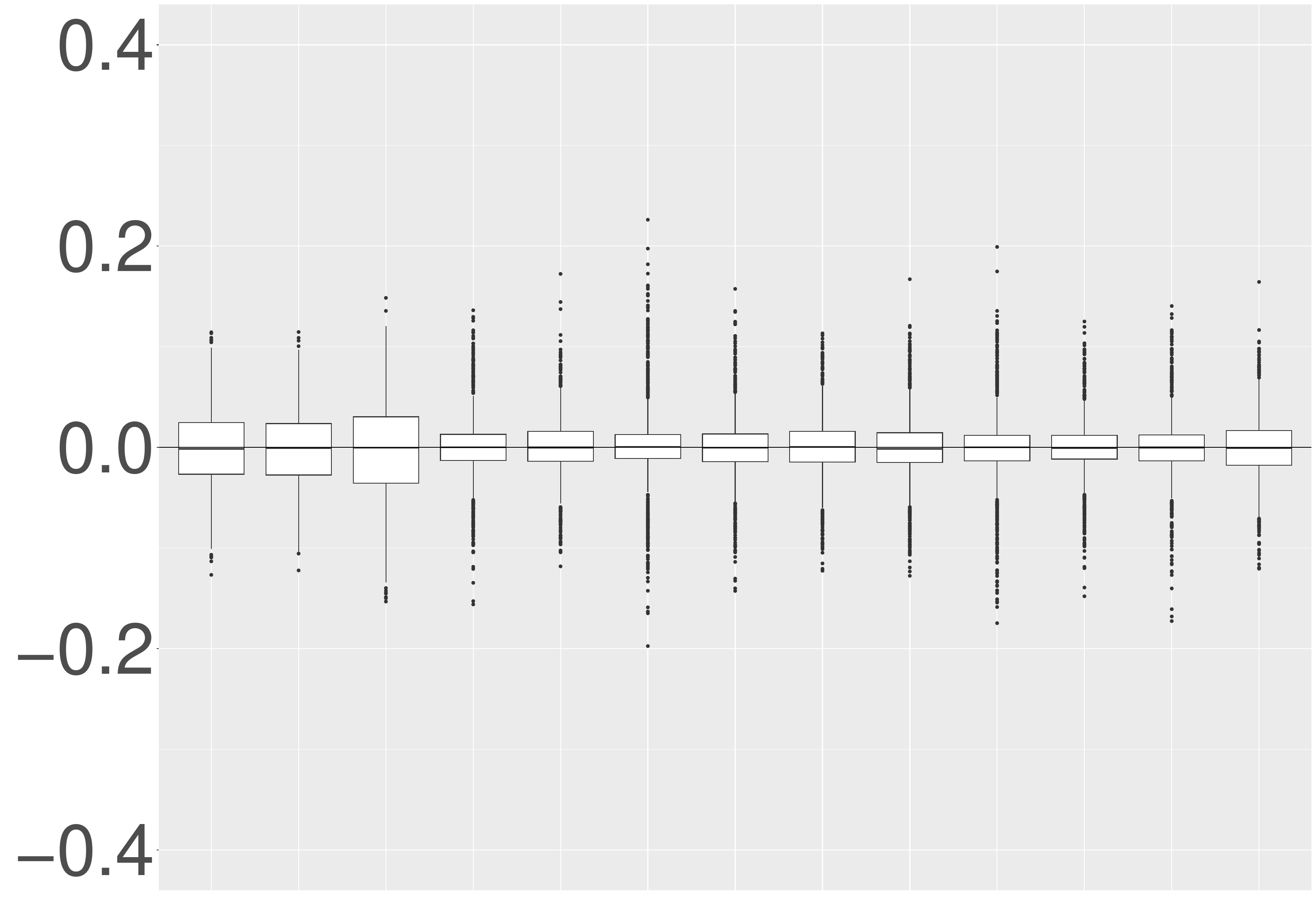}
\end{subfigure}
\begin{subfigure}[b]{.30\textwidth}
    \centering
    \includegraphics[width=\textwidth]{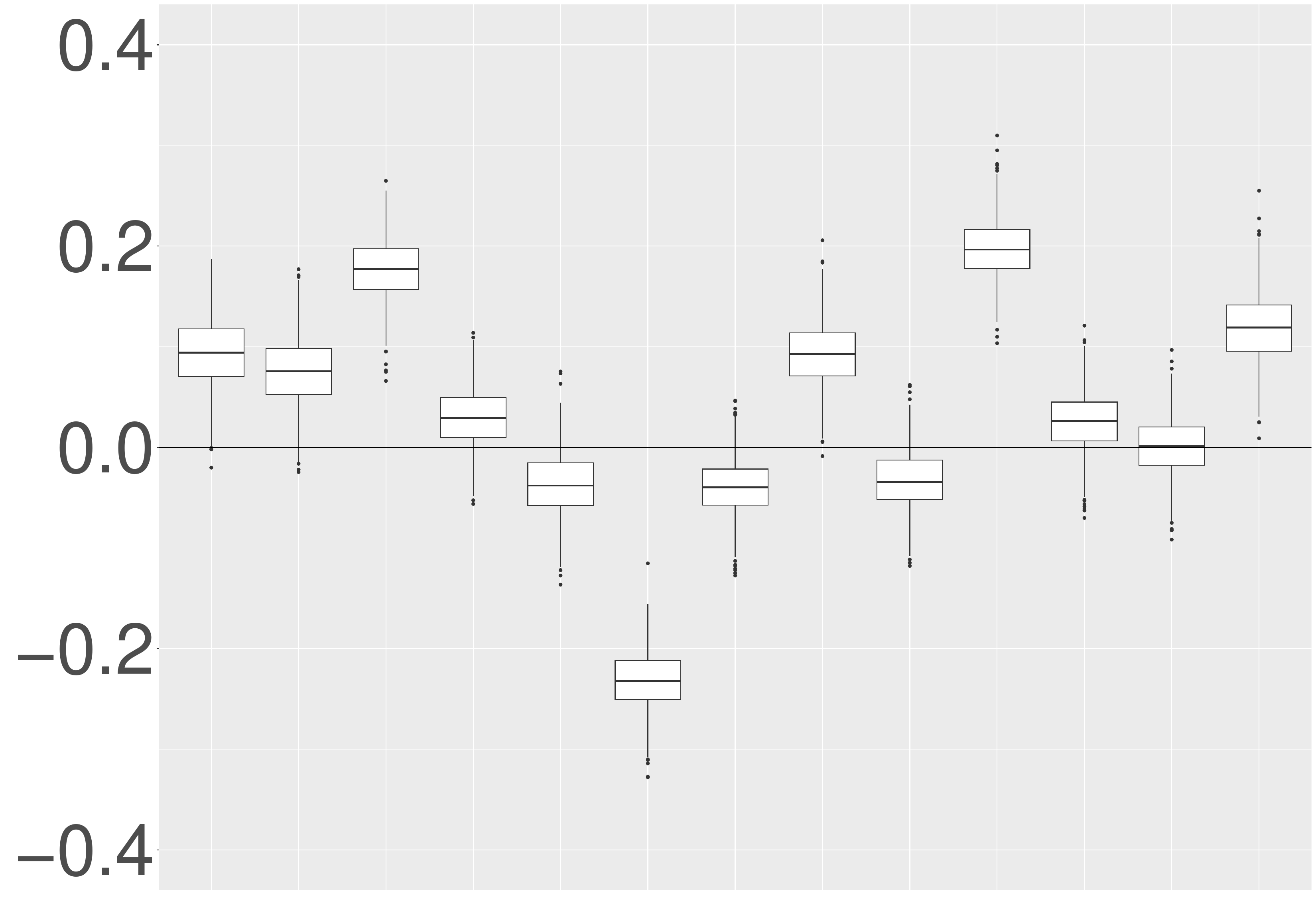}
\end{subfigure}

\makebox[10pt]{\raisebox{50pt}{\rotatebox[origin=c]{90}{\small{s=6}}}}%
\begin{subfigure}[b]{.30\textwidth}
    \centering
    \includegraphics[width=\textwidth]{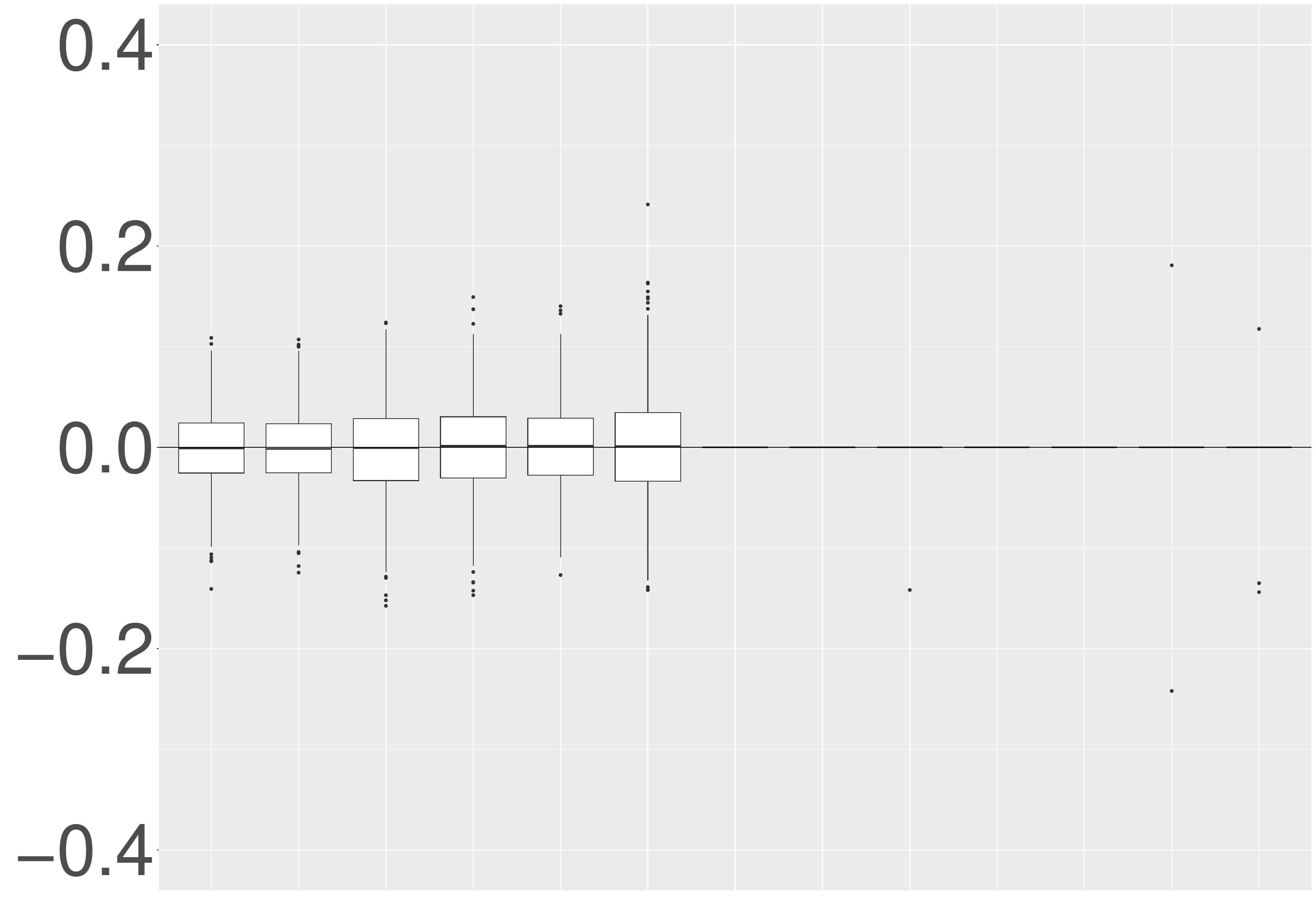}
\end{subfigure}
\begin{subfigure}[b]{.30\textwidth}
    \centering
    \includegraphics[width=\textwidth]{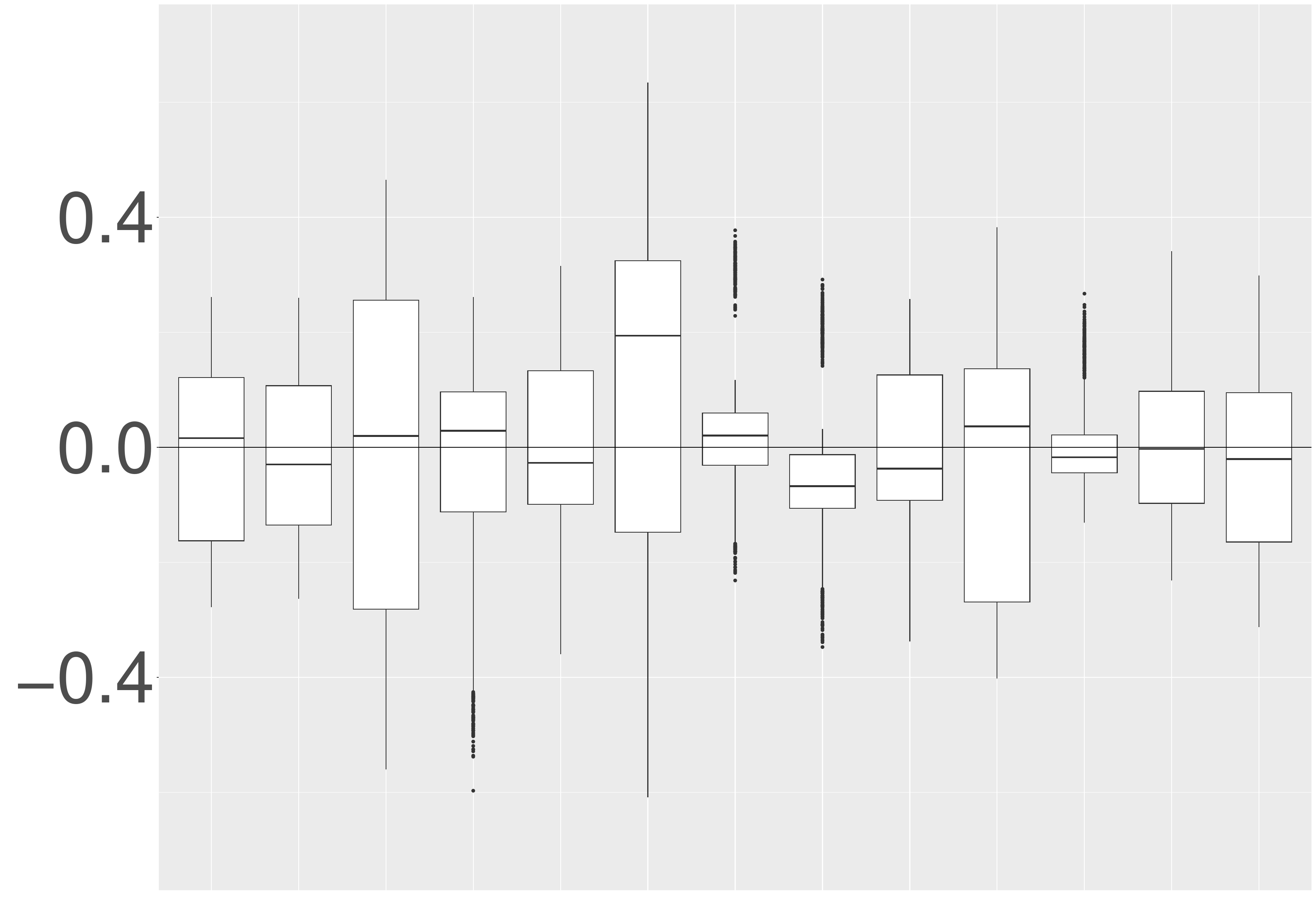}
\end{subfigure}
\begin{subfigure}[b]{.30\textwidth}
    \centering
    \includegraphics[width=\textwidth]{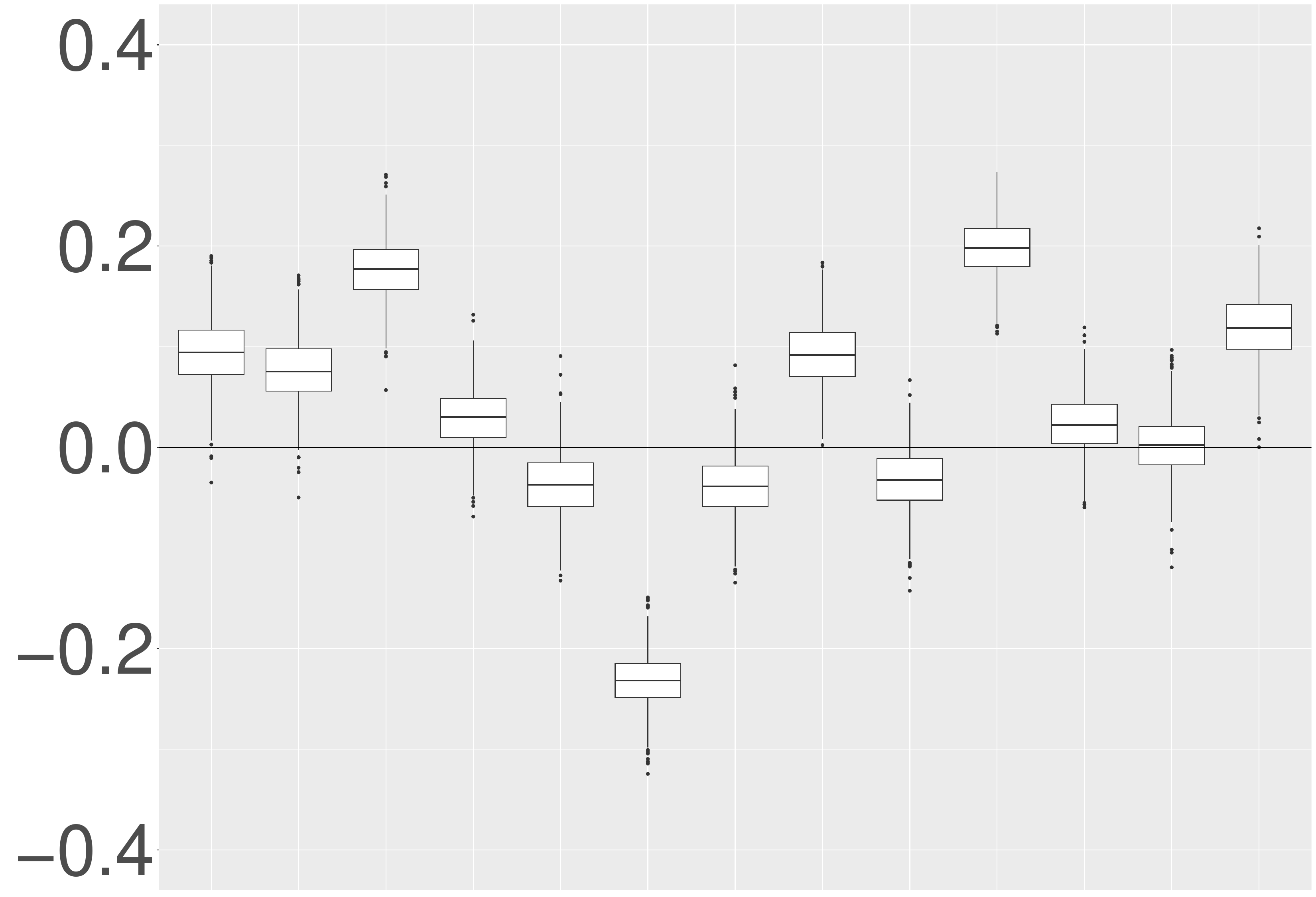}
\end{subfigure}

\makebox[10pt]{\raisebox{78pt}{\rotatebox[origin=c]{90}{\small{s=8}}}}%
\begin{subfigure}[b]{.30\textwidth}
    \centering
    \includegraphics[width=\textwidth]{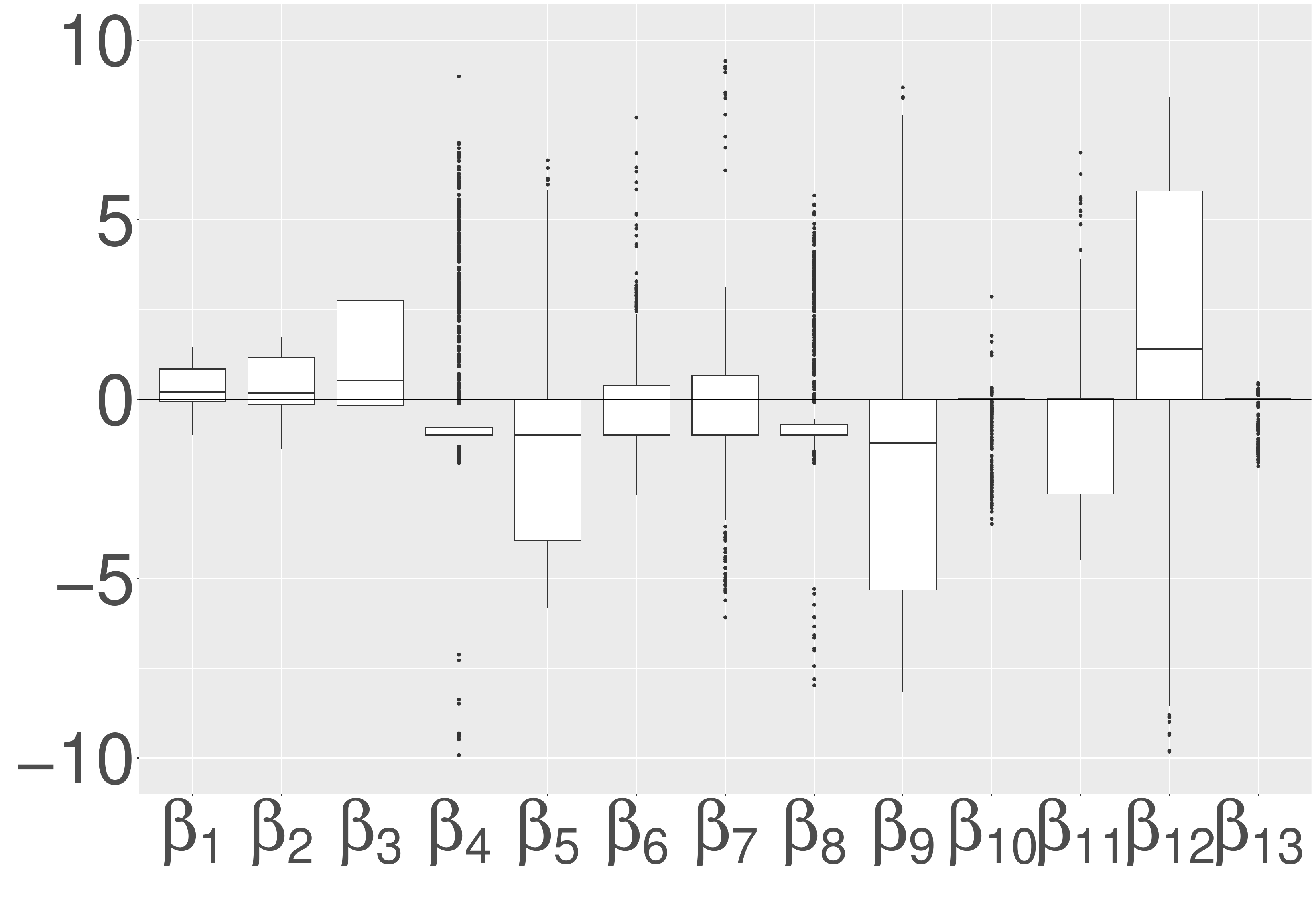}
    \caption{Spar}
\end{subfigure}
\begin{subfigure}[b]{.30\textwidth}
    \centering
    \includegraphics[width=\textwidth]{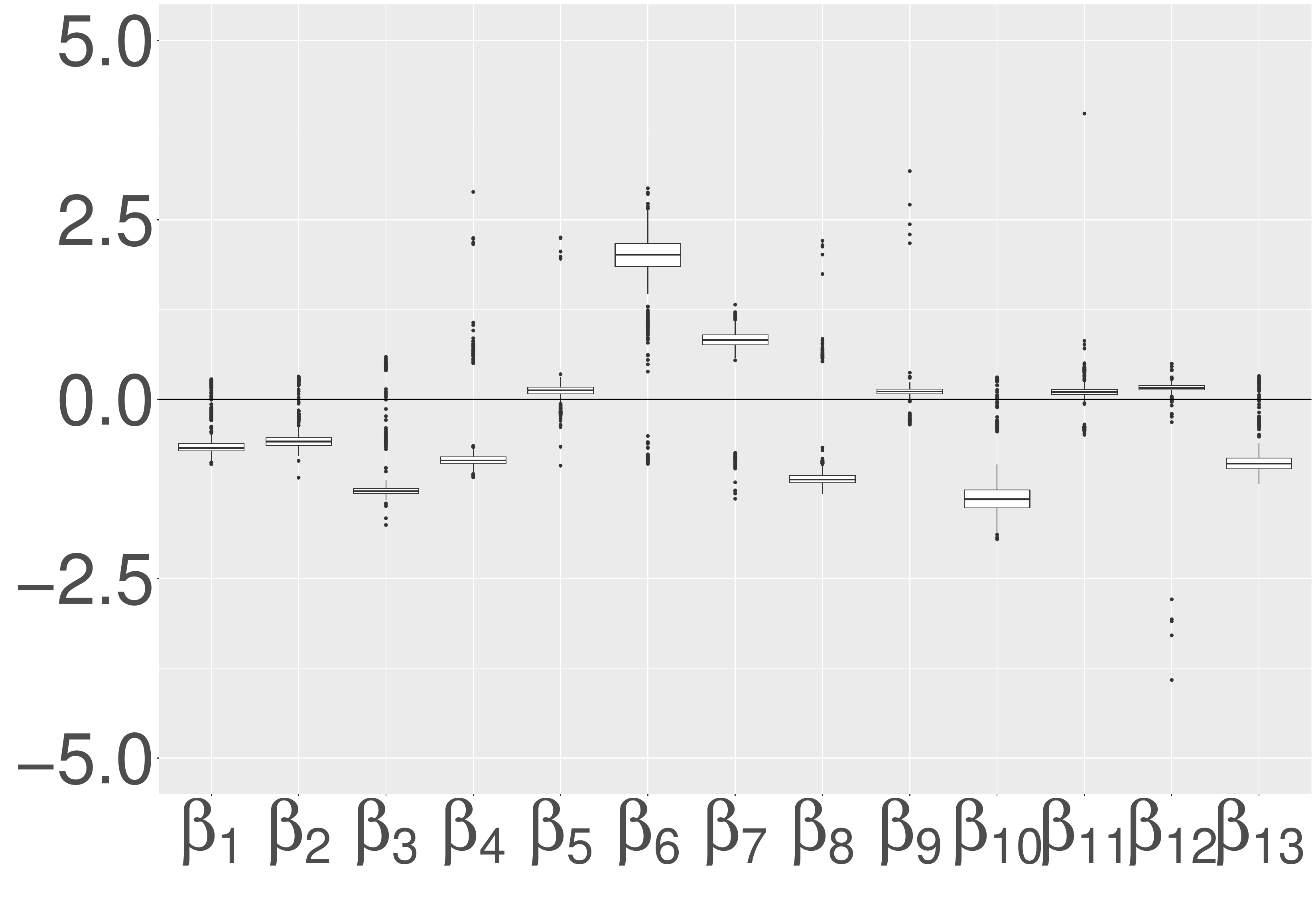}
    \caption{Null}
\end{subfigure}
\begin{subfigure}[b]{.30\textwidth}
    \centering
    \includegraphics[width=\textwidth]{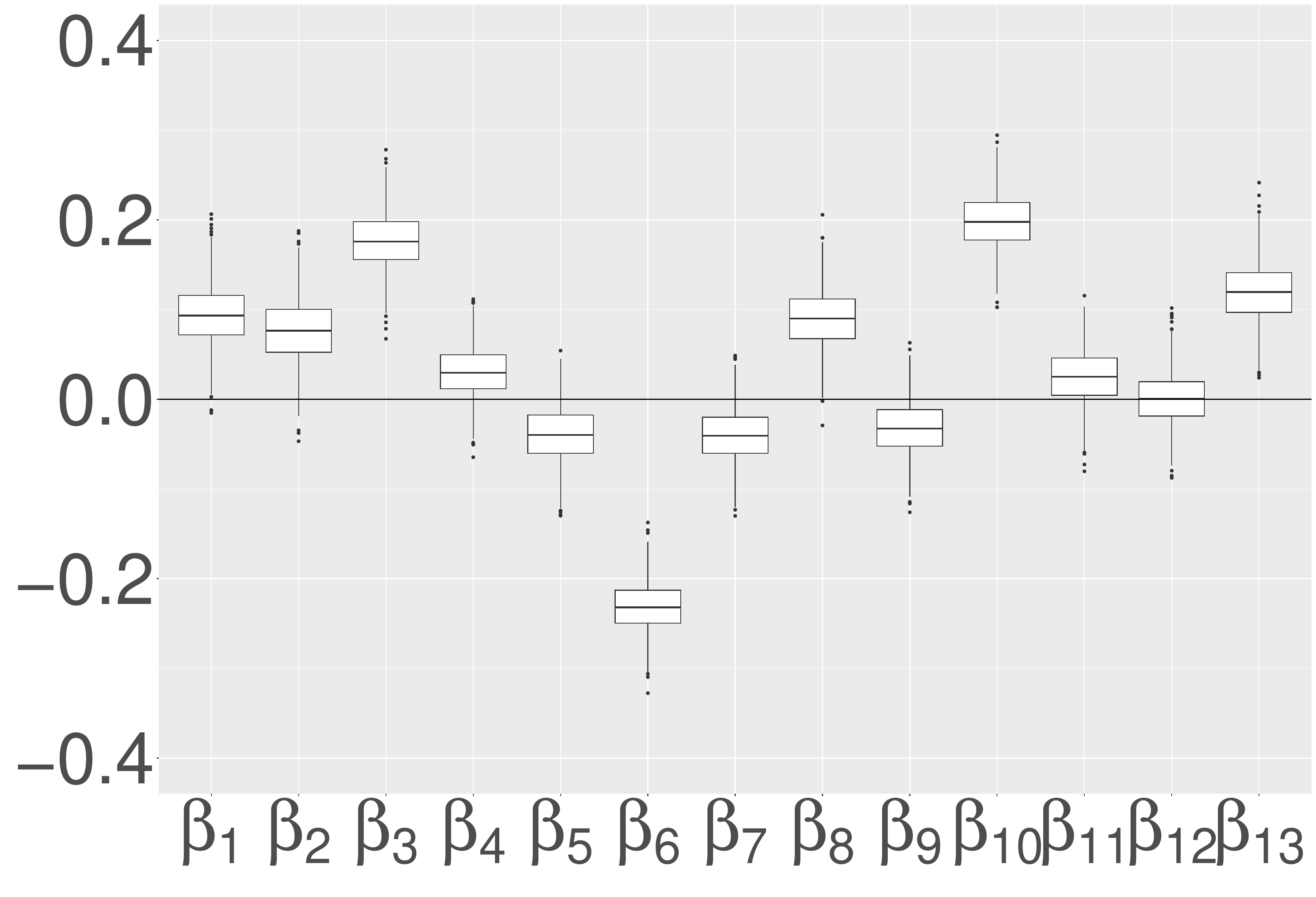}
    \caption{OLS}
\end{subfigure}
\caption{Boxplots of the estimation bias for Spar, Null, and OLS when the sparsity level $s = 3,6,8$. Here, $s$ represents the number of non-zero entries in the causal effects $\beta$. The horizontal line at zero indicates no bias. The Spar method exhibits minimal bias centered at 0 for both $s=3$ and $s=6$, whereas the Null method has little bias only at $s=3$.}
\label{fig:box}
\vspace{-3mm}
\end{figure}

Figure~\ref{fig:box} shows the estimation bias of individual elements within $\beta$. The estimation bias of Spar is centered around 0 when $s = 3$ and $6$, while Null has minimal bias only when $s=3$. Note that when $s = 3, 6$, the estimation bias of $\beta_i$ when $\beta_i = 0$ for the proposed Spar method is zero, except for a few outliers. This illustrates that the proposed Spar estimator is sparse and effectively selects the active treatments that have non-zero causal effects on the outcome. When the number of non-zero entries in $\beta$ increases (e.g., $s=8$), both the Spar and Null methods exhibit noticeable estimation bias. The OLS estimator is biased at every sparsity level due to unmeasured confounding. In summary, although the performance of both the Spar and Null methods deteriorates when the sparsity assumptions are severely violated, the Spar method exhibits a less stringent sparsity constraint compared to the Null method, both theoretically and empirically.


\begin{remark}
    In our estimation method, we introduce an extra step in which we identify a set \(\mathcal{I}\) consisting of the zero indices from the initial estimate \(\widehat{\beta}^{mip}\), resulting in \(\xi_i = \gamma_i \delta\) for \(i \in \mathcal{I}\). Based on our numerical results, \(\mathcal{I}\) typically identifies the nonzero indices accurately, allowing for unbiased and asymptotically normal estimates of \(\delta\) and \(\beta\) through simple linear regression. This additional step mitigates the bias in \(\widehat{\delta}\) from the mixed-integer programming approach, thereby improving overall estimation accuracy.
\end{remark}

\subsection{High dimensional case}
\label{sec:highd}

In this part, we consider the case where the number of treatments $p$ is larger than the number of observations $n$. We set $q = 3$, and define $\beta = (1, 1, 1, 1, 1, 0, \ldots, 0)^T$, where the first 5 entries of $\beta$ are set to 1, and the remaining entries are all 0. In each Monte-Carlo simulation, we generate each element in $\alpha \in \mathcal{R}^{p\times q}$ and $\delta \in \mathcal{R}^{q}$ independently and identically distributed from $Uniform(-1,1)$. We simulate the latent confounders $U$ from $N(0, I_q)$. The random errors are generated from $\epsilon_x \sim N(0, 2I_p)$ and $\epsilon_y \sim N(0,1)$. We set the sample size $n=300$ and vary $p$ from 300 to 1000.

As a baseline model, we implement the Lasso \citep{Lasso} using the \textit{glmnet} function in {\tt R} and tune the hyperparameter using 10-fold cross-validation. To adapt the null treatment method in \cite{Miao} to the high-dimensional setting, we first obtain estimates for $\widehat{\xi}$ and $\widehat{\gamma}$ using the same method as employed in the Spar method, and then apply the least median squares method as described in \cite{Miao}. Under the above simulation settings, the sparsity assumptions are satisfied for both the Spar and Null methods. We also consider the deconfounder method in \cite{deconfounder}, denoted as Deconf, for comparison. Specifically, we apply probabilistic principal component analysis (PPCA) with 50 latent dimensions to the treatments $X$ to obtain a substitute for the unobserved confounders. We then utilize Lasso regression on the augmented data to estimate $\beta$. The following experimental results are based on 100 Monte-Carlo simulations.

We first assume that $q=3$ is correctly specified, and later in this section, we will explore the performance of Spar when we misspecify the number of latent confounders.

\begin{figure}[htbp!]
\centering
\begin{subfigure}[b]{.45\textwidth}
    \centering
    \includegraphics[width=\textwidth]{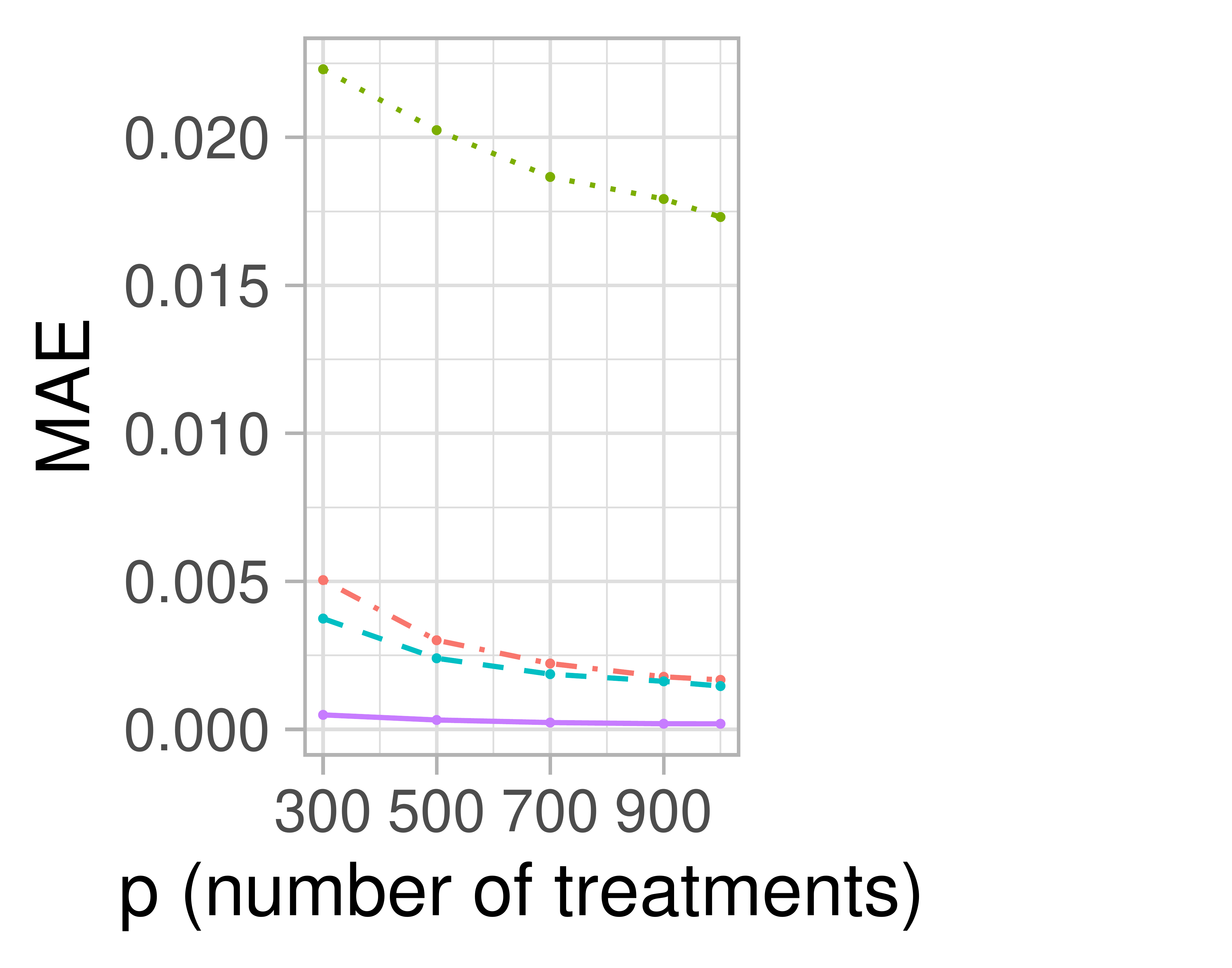}
\end{subfigure}\hspace{-2em}%
\begin{subfigure}[b]{.45\textwidth}
    \centering
    \includegraphics[width=\textwidth]{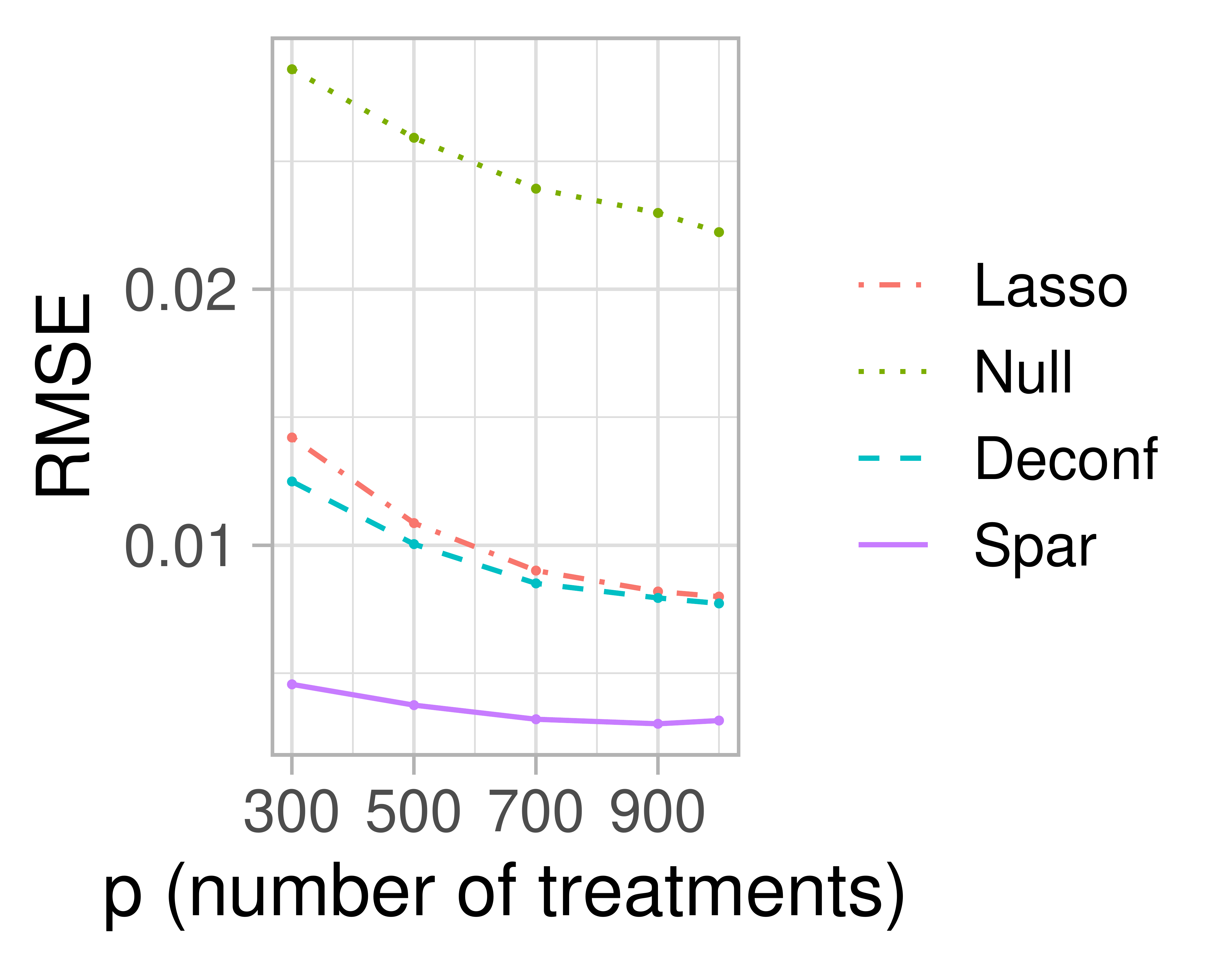}
\end{subfigure}
\caption{The MAE and RMSE of Spar, Null, Deconf and OLS methods under different number of treatments when $n=300$. The sparsity assumption is met for both the Spar and Null methods.}
\label{fig:errorhighd}
\end{figure}

Figure~\ref{fig:errorhighd} presents the MAE and RMSE for the Spar, Null, Deconf, and OLS methods as the number of treatments $p$ increases. The Spar method consistently demonstrates the smallest MAE and RMSE among all methods across different values of $p$. The Deconf method shows a slight improvement over the naive lasso approach. The Null method has the largest estimation error, as the causal effect $\beta$ in our simulation is sparse, and all methods except for the Null yield sparse estimators.

One notable feature of our method is its capacity to select zero entries within $\beta$ as part of the estimation process. Lasso also provides sparse estimates of $\beta$. We compare the ability of both methods to correctly identify the non-zero causal effects and report their true positive rate (TPR) and false positive rate (FPR) {in Table \ref{tab:FPR}}. Both methods are effective in accurately identifying active treatments that have non-zero causal effects on the outcome, with TPR close to 1. However, the Lasso method has a noticeable probability of incorrectly identifying zero $\beta$ values as non-zero. In contrast, our method can select treatments that causally influence the outcome with high precision while ruling out the treatments with no causal effect on the outcome.


\begin{table}[htbp!]
\centering
\caption{The true positive rate (TPR) and the false positive rate (FPR) for Lasso and Spar methods under different numbers of treatments ($p$).}
\begin{tabular}{@{}lllllll@{}}
\toprule
\multicolumn{2}{l}{$p$}      & 300                   & 500                   & 700                   & 900                   & 1000                  \\ \midrule
\multirow{2}{*}{TPR} & Lasso & 1.000                 & 1.000                 & 1.000                 & 1.000                 & 1.000                 \\
                     & Spar  & 1.000                 & 1.000                 & 1.000                 & 1.000                 & 1.000                 \\ \midrule
\multirow{2}{*}{FPR} & Lasso & 0.222                 & 0.143                 & 0.114                 & 0.092                 & 0.088                 \\
                     & Spar  & 6.780$\times 10^{-6}$ & 6.061$\times 10^{-6}$ & 4.317$\times 10^{-6}$ & 7.449$\times 10^{-6}$ & 3.350$\times 10^{-6}$ \\ \bottomrule
\end{tabular}
\label{tab:FPR}
\end{table}

\begin{figure}[htbp!]
\centering
\begin{subfigure}[b]{.45\textwidth}
    \centering
    \includegraphics[width=\textwidth]{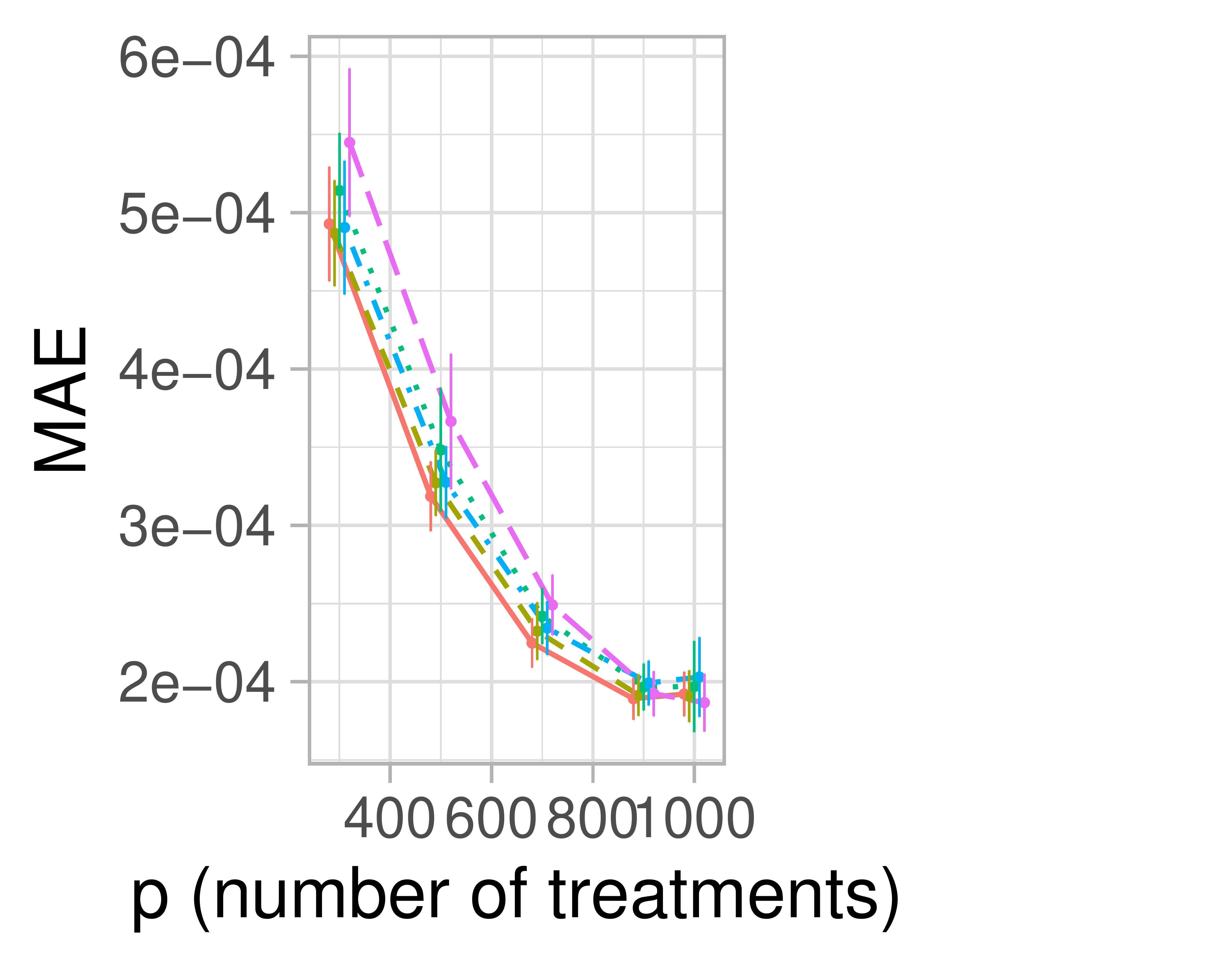}
\end{subfigure}\hspace{-3em}%
\begin{subfigure}[b]{.45\textwidth}
    \centering
    \includegraphics[width=\textwidth]{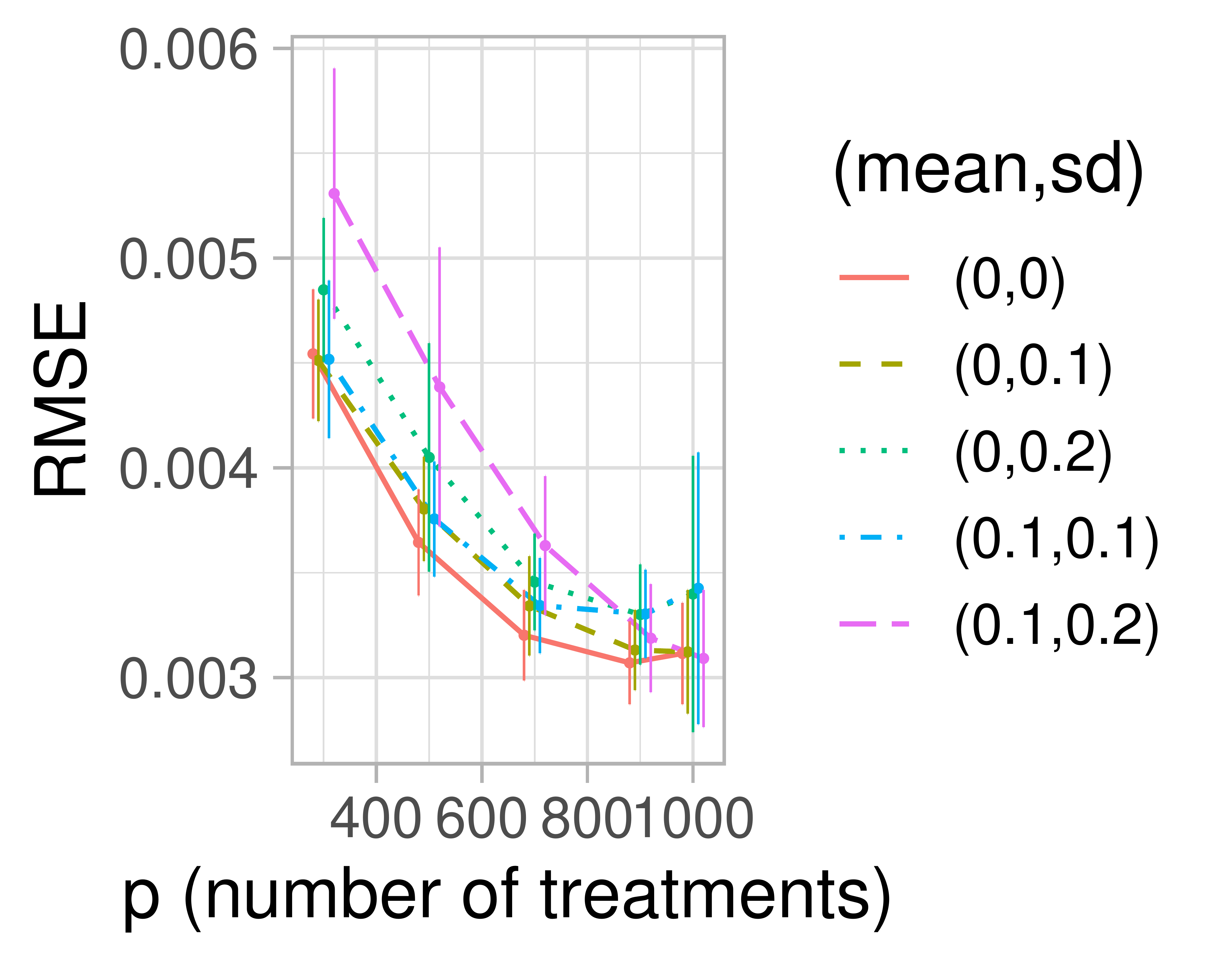}
\end{subfigure}
\caption{The MAE and RMSE for the Spar method under different non-diagonal sparse matrix settings of $\Sigma_{\epsilon_x}$, with 95\% confidence intervals. The mean and standard deviation of the off-diagonal entries are specified in (mean, sd). (0,0) represents the setting where $\Sigma_{\epsilon_x}$ is diagonal.}
\label{fig:nondiag}
\end{figure}

The Assumption $A1$ states that the noises of treatments are uncorrelated to identify $\alpha$ via factor analysis. This assumption is not required in high-dimensional settings where $p>n$, as the PCA estimator of the factor loading utilized in our method allows for the existence of sparse correlations between $\epsilon_x$ \citep{PCAhidimfact}. Following the sparse covariance matrix generation approach in \cite{POET}, we evaluate the robustness of the Spar method when the conditional independence of treatments given the latent confounders is violated. We consider four different combinations of the mean and standard deviation of the off-diagonal entries of $\Sigma_{\epsilon_x}$: (0,0.1), (0,0.2), (0.1,0.1), (0.1,0.2). Figure~\ref{fig:nondiag} presents the MAE and RMSE, along with their 95\% confidence intervals across different mean and standard deviation configurations. The setting (0,0) represents the scenario when $\Sigma_{\epsilon_x}$ is diagonal. While the estimation errors are the smallest when Assumption $A1$ holds, the confidence intervals for the estimation errors in both diagonal and non-diagonal settings are highly overlapped, especially as $p$ increases. This shows that in the high-dimensional setting where the number of treatments is large, the Spar method could accommodate the case when the conditional independence of treatments $X$ given the unmeasured confounders $U$ is violated, and the noises of $X$ have sparse non-diagonal covariance matrix.


\begin{figure}[htbp!]
\centering
\begin{subfigure}[b]{.45\textwidth}
    \centering
    \includegraphics[width=\textwidth]{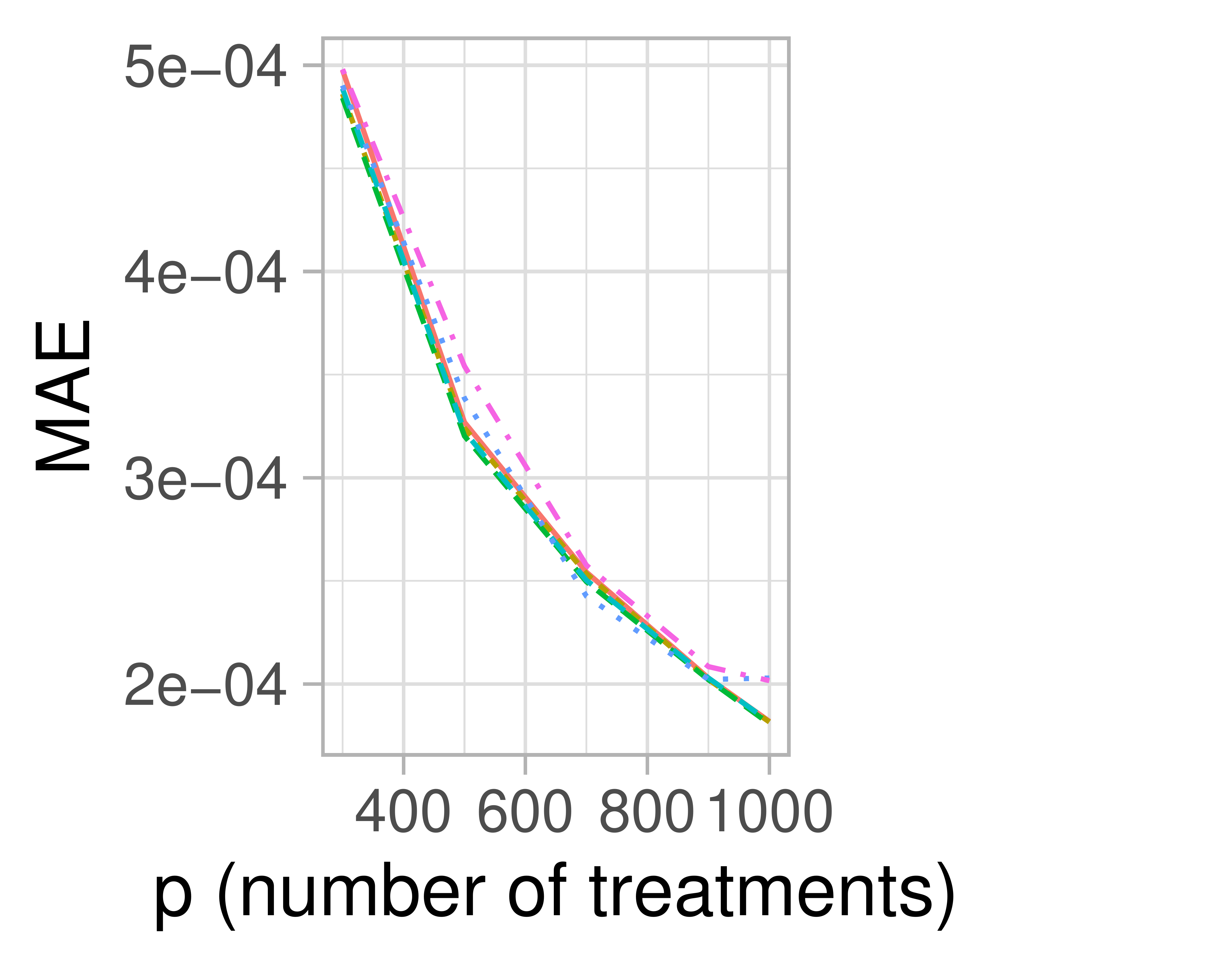}
\end{subfigure}\hspace{-2em}%
\begin{subfigure}[b]{.45\textwidth}
    \centering
    \includegraphics[width=\textwidth]{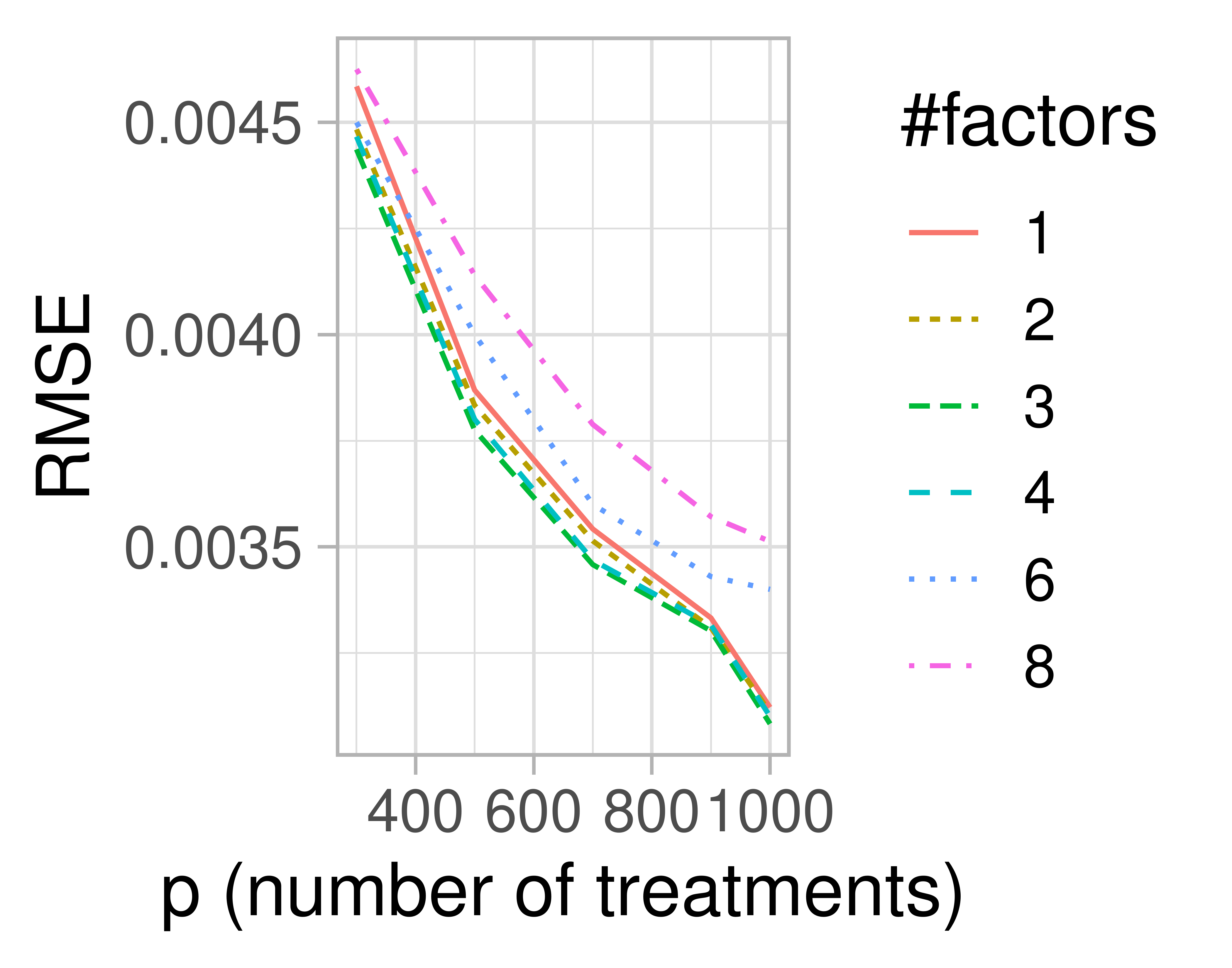}
\end{subfigure}
\caption{The MAE and RMSE for Spar method under different specifications of number of unmeasured confounders. $q=3$ is the true number of confounders.}
\label{fig:nfact}
\end{figure}

In the above simulations, we assume that $q$ is correctly specified. However, in practice, it is not always the case, and we may need to estimate $q$ from data. Therefore, we assess the robustness of the Spar method under different specifications of $q$. As shown in Figure~\ref{fig:nfact}, the estimation error for $q=3$ is the smallest among different specifications of the dimension of confounders. When the specified $q$ is not significantly different from the true value (e.g., when $q=2,4$), the estimation errors show little increase. This suggests that the performance of the Spar method is robust to a slight deviation from the true number of latent confounders. The estimation errors become larger as the difference between the misspecified dimension of confounders and the truth increases. In practice, it is recommended to assess the robustness of estimation by varying the specified number of latent confounders in data analyses.

\begin{figure}[htbp!]
\centering
\includegraphics[width=4.5in]{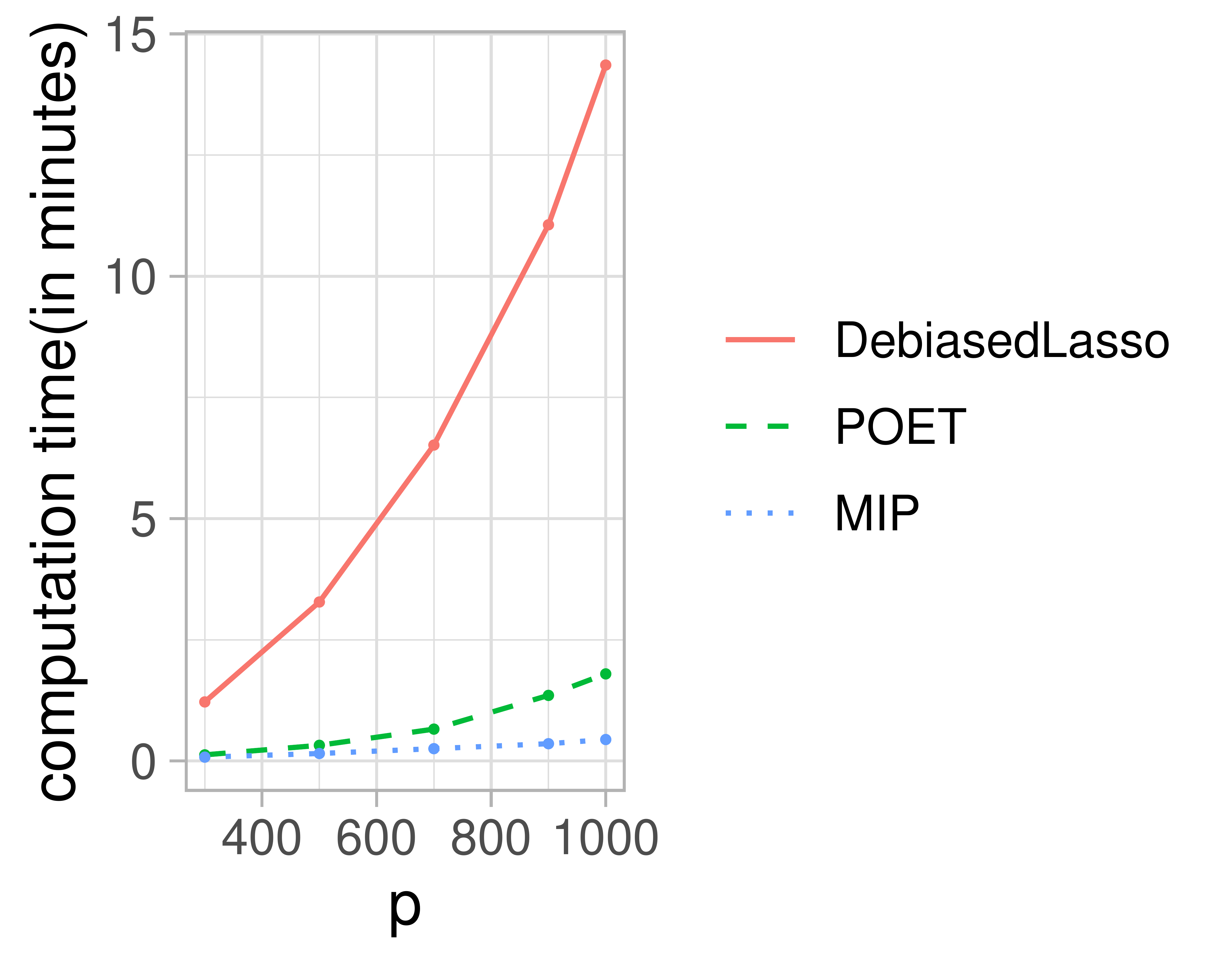}
\caption{The computation time (in min) for de-biased lasso, POET, and mixed integer programming (MIP) steps in the Spar method.} 
\label{fig:time}
\end{figure}

In high-dimensional settings, computational efficiency is a major concern. Three steps in our approach are potentially time-consuming: the de-biased lasso estimation of $\xi$, the POET estimation of the covariance matrix of $X$, and the mixed-integer programming in (\ref{equ:mip}). Figure~\ref{fig:time} shows the average computation time for a single Monte Carlo simulation. The experiments were conducted on a 2021 14$''$ MacBook Pro (10-core M1 Pro, 32GB unified memory). The most time-consuming step is the de-biased lasso estimation, with the required time increasing as the number of treatments grows. The mixed-integer programming step requires the least time. Although MIP problems are NP-hard in the worst case, some problems can be solved efficiently in practice depending on the structure of the constraints and objectives. The computation time of MIP in our method is relatively short and does not significantly increase as $p$ increases. 



\section{Synthetic GWAS Data}
\label{sec:gwas}

In genome-wide association studies (GWAS), it is of interest to investigate the causal relationships between genetic variations and traits. Genetic data contains high-dimensional single-nucleotide polymorphisms (SNPs) serving as multiple treatments. However, only a subset of these SNPs has a causal effect on the studied trait. Additionally, various latent confounders, such as population structure and lifestyle variables, correlate with both the treatments and the outcome. In this section, we evaluate the performance of our method using a synthetic GWAS dataset previously utilized in \cite{deconfounder} and \cite{deconfnaive}. Below are the details of how this synthetic dataset is generated.

Let $\mathbf{a}_i \in \mathcal{R}^p$ represent the SNP values for individual $i$, and $y_i$ denote the real-valued trait for individual $i$. The genotypes for $n$ individuals are structured in the matrix $A = (\mathbf{a}_1, \mathbf{a}_2, \ldots, \mathbf{a}_n)^T \in \mathcal{R}^{n \times p}$. In accordance with \cite{deconfounder}, we simulate $A$ using $A_{ij} \sim \text{Binomial}(2, P_{ij})$, where $P \in \mathcal{R}^{n \times p}$ represents the allele frequencies. We set $P = S \Gamma$, with $\Gamma \in \mathcal{R}^{d \times p}$ and $S \in \mathcal{R}^{n \times d}$, where $d < p$. Matrix $S$ encodes the genetic population structure, and $\Gamma$ maps how this structure influences the allele frequencies of each SNP. Following \cite{deconfounder} and \cite{deconfnaive}, we employ three realistic models—Balding-Nichols Model (BN), Pritchard-Stephens-Donnelly model (PSD), and Spatial model—to generate $S$ and $\Gamma$. Detailed configurations for these models are provided in Section C of the supplementary material. For each scenario, we generate $p = 1000$ SNPs for $n = 1000$ individuals. The generation of genotypes $\textbf{a}_i$ closely mimics real-world structures. Specifically, the BN and PSD models utilize the SNP structure from the HapMap dataset, while the Spatial model replicates a smooth spatial mixing of populations. However, our Spar estimator is based on the linear factor model (\ref{equ:model}). We assess the performance of our method on synthetic datasets to assess its performance under potential model misspecifications.


We simulate the trait $y_i$ for each individual under a linear model consisting causal signals, confounders, and random effects,
\begin{equation}
    y_i = \sum_{j=1}^p \beta_j a_{ij} + \lambda_i + \epsilon_i.
\end{equation}
Analogous to the sparse causal effects in the real world, we set the first 1\% of the $p$ genes to be the true causal treatments ($\beta = 0.5$), with the remaining $\beta$ values set to zero.  The confounders $\lambda_i$ and the random effects $\epsilon_i$ are sampled based on the latent population structure $S$. We first perform K-means clustering on the rows of $S$ with $K = 3$, which assigns each individual $i$ to one of the three clusters $\mathcal{S}_1$, $\mathcal{S}_2$, $\mathcal{S}_3$. We set $\lambda_i = k$ for $i \in \mathcal{S}_k$. Let $\tau_1^2,\tau_2^2,\tau_3^2 \overset{iid}{\sim} InvGamma(3,1)$, and draw $\epsilon_i \sim N(0,\tau_k^2)$ for each individual $i$ in class $k$.


Furthermore, we impose different signal-to-noise ratio (SNR) settings, where causal signals, confounders, and random effects contribute different proportions of the variance in the trait $y_i$. Denote $v_{gene}$, $v_{conf}$, and $v_{noise}$ as the proportions for the three components, respectively, with the constraint $v_{gene} + v_{conf} + v_{noise} = 1$. The SNR is defined as $(v_{gene}+v_{conf}) / v_{noise}$, where $v_{gene} = v_{conf}$. We consider SNR values from the set \{0.1, 0.5, 0.7, 1.0, 3.0, 5.0, 7.0, 9.0, 11.0, 13.0\}. We adjust $\lambda_i$ and $\epsilon_i$ to achieve the desired signal-to-noise ratio:
$$
\lambda_{i} \leftarrow\left[\frac{s . d .\left\{\sum_{j=1}^{p} \beta_{j} a_{i j}\right\}_{i=1}^{n}}{\sqrt{v_{\text {gene }}}}\right]\left[\frac{\sqrt{v_{c o n f}}}{s . d .\left\{\lambda_{i}\right\}_{i=1}^{n}}\right] \lambda_{i}, 
$$
$$
\epsilon_{i} \leftarrow\left[\frac{s . d .\left\{\sum_{j=1}^{p} \beta_{j} a_{i j}\right\}_{i=1}^{n}}{\sqrt{v_{\text {gene }}}}\right]\left[\frac{\sqrt{v_{n o i s e}}}{s . d .\left\{\epsilon_{i}\right\}_{i=1}^{n}}\right] \epsilon_{i}.
$$

In \cite{deconfnaive}, a comparison is drawn between the naive ridge regression and the deconfounder method for the synthetic GWAS data. The results indicate that applying the deconfounder method in \cite{deconfounder} to ridge regression provides only marginal improvements over the naive ridge estimator. In addition to the Lasso, Null, Deconf, and Spar methods, we also evaluate the performance of both the naive ridge regression and the deconfounder method based on ridge regression in this section. Ridge regression is implemented using the \textit{glmnet} function in {\tt R}. The Deconf(ridge) method incorporates SNP values $a_i$ and the estimated substitute confounders as its regressors. Experimentation involves 100 Monte-Carlo simulations across three genotype models (BN, PSD, and Spatial) and various Signal-to-Noise Ratio (SNR) settings.



\begin{figure}[htbp!]
\centering
\includegraphics[width=1\textwidth]{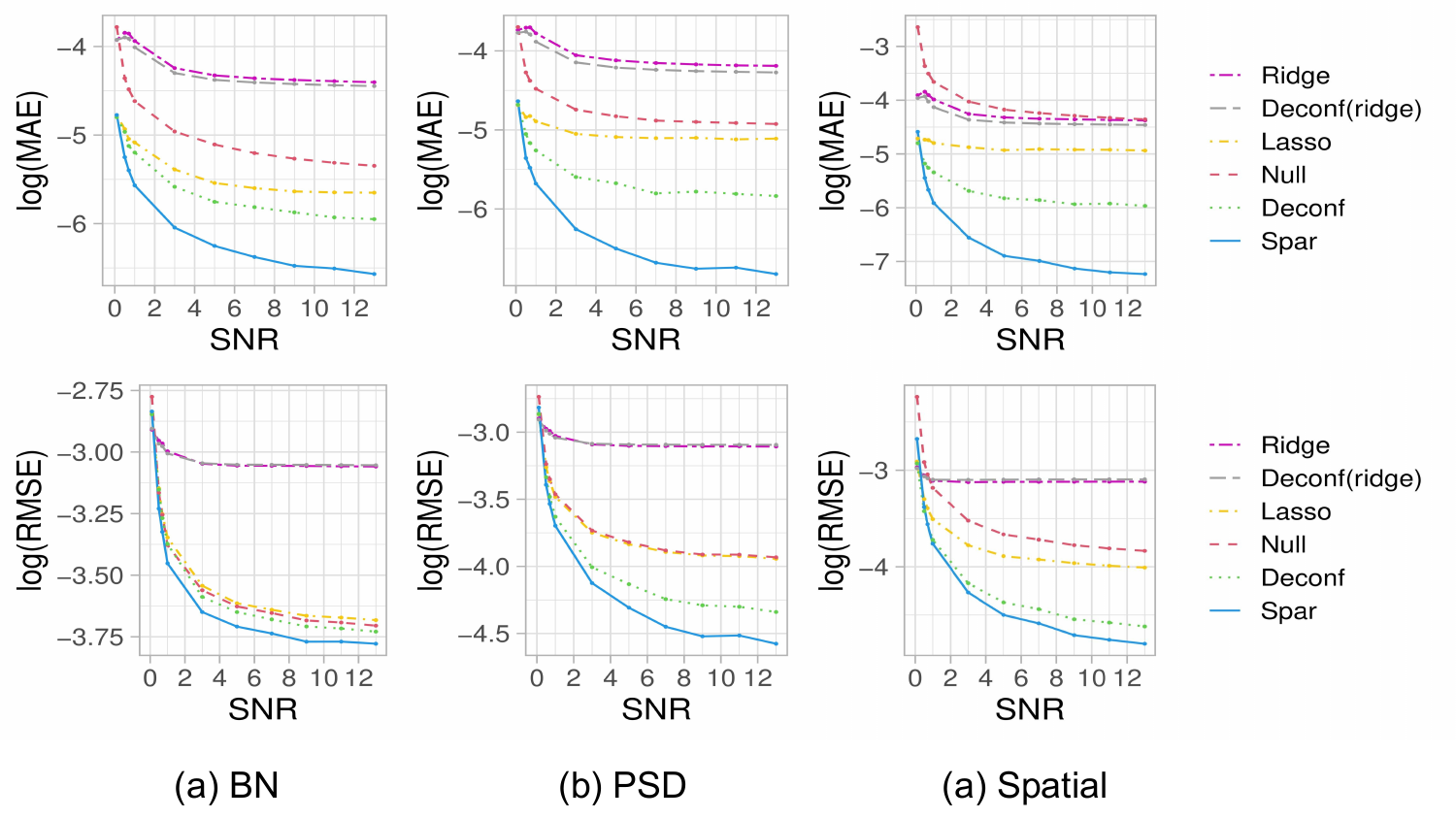}
\caption{MAE and RMSE of the method Spar, Null, Deconf, Deconf(ridge), Lasso, and Ridge for the three models (BN, PSD, and Spatial) under different signal-to-noise rates.}
\label{fig:gwas}
\end{figure}

As shown in Figure \ref{fig:gwas}, our method demonstrates superior performance compared to the Deconf, Null, Ridge, and Lasso estimators when the SNR is not too small (e.g., SNR $> 1$). Under low SNR conditions (SNR $<1$), the MAE and RMSE of the Lasso and Spar methods are similar, as it becomes challenging to identify the true causal effects when the noise variance is larger than those of the causal signals and confounders. As SNR increases, the MAE and RMSE of all six methods first decrease, and then converge to a certain value. The Lasso, Deconf, and Spar methods, which give sparse estimators of $\beta$, consistently outperform the Null, Ridge, and Deconf(ridge) methods, since we have sparse causal effects $\beta$. Although the GWAS datasets deviate from the factor model of the treatment $X$ in (\ref{equ:model}), our approach maintains robust performance. It is also worth mentioning that the estimation errors of Deconf(ridge) under the three models are close to or even higher than those of ridge regression. This observation aligns with the findings in \cite{deconfnaive} that the deconfounder shows marginal improvements or even worse performance than the naive estimator.


The Assumption $A1$ requires that the causal effects of the majority of the treatments should be zero. In our previous simulations, the Assumption $A1$ is met since we assumed that the causal effects of the last 99\% genes were precisely zero. However, it may not fully capture the complexity of real-world scenarios where many treatments may have weak but non-zero causal effects. We account for such occasions in Section D of the supplementary material. Specifically, we assign a uniform distribution to the last 99\% causal effects $\beta \sim Uniform$(-0.05,0.05), introducing perturbations to these causal effects. Our findings show that the proposed Spar method is robust even when the sparsity assumption is not strictly met.



\section{Discussion}
\label{sec:discuss}

In this paper, we introduce a novel method for causal effect estimation in a multi-treatment setting. Our major assumption is that the causal effect $\beta$ is sparse, with more than $(p+q)/2$ zero entries, where $p$ is the number of treatments, and $q$ is the number of unmeasured confounders. In comparison to the estimation method introduced by \cite{Miao} for a similar multi-treatment scenario, our proposed estimation method relies on a more relaxed sparsity assumption. Moreover, even when both sparsity conditions are met, our approach achieves sparse estimation by minimizing the $\ell_0$ norm, resulting in lower mean absolute error (MAE) and root mean squared error (RMSE) compared to alternative methods.



We assume a linear outcome model and a factor model for the treatments. Exploring the causal identification and estimation techniques for cases with non-linear outcome models, or when treatments and confounders do not adhere to a factor model, remains a subject for future investigation. Nonetheless, our analysis of synthetic genome-wide association study (GWAS) data demonstrates that the proposed method generalizes well even when treatments and unmeasured confounders deviate from a factor model.

A prerequisite for our estimation procedure is that the dimension of unmeasured confounders $U$ is known. Thus, it is necessary to first determine the dimension of latent confounders. Existing methods for identifying the number of latent confounders in factor models, such as those proposed by \cite{nfact} and \cite{nfactOnatski}, can be employed for this purpose. We recommend a sensitivity analysis by varying the number of latent confounders during data analysis to assess the robustness of the estimator.


 
\bigskip
\begin{center}
{\large\bf SUPPLEMENTARY MATERIALS}
\end{center}

\begin{description}

\item[Title:] Supplement to ``Simultaneous Estimation of Multiple Treatment Effects from Observational Studies" (.pdf file)

\item[R code:] R code to reproduce the simulation results in the article. (.zip file)



\end{description}

\bibliographystyle{chicago}

\bibliography{main}

\end{document}


\def\spacingset#1{\renewcommand{\baselinestretch}%
{#1}\small\normalsize} \spacingset{1}

\spacingset{1.75}

\if0\blind
{
  \title{\bf Supplementary Materials for ``Simultaneous Estimation of Multiple Treatment Effects from Observational Studies"}
  \author{Xiaochuan Shi, Dehan Kong, and Linbo Wang\thanks{
    The authors gratefully acknowledge funding from the Natural Sciences and Engineering Research Council of Canada and Canadian Statistical Sciences Institute for supporting this research.}\hspace{.2cm}\\
    Department of Statistical Sciences, University of Toronto}
    \date{}
  \maketitle
} \fi

\if1\blind
{
  \bigskip
  \bigskip
  \bigskip
  \begin{center}
    {\LARGE\bf Supplementary Materials for ``Simultaneous Estimation of Multiple Treatment Effects from Observational Studies"}
      \date{}
\end{center}
  \medskip
} \fi

\bigskip

\appendix

\section{Proof of Theorem 1}
\label{supp:proof}

The first equation in the assumed linear structural model (1) is a classical factor model. According to \cite{factana}, the loading matrix $\alpha$ is uniquely identified up to a rotation under the assumptions $A1$ and $A2$. Let $R$ be an arbitrary rotation matrix, and $\alpha_1 = \alpha R$, $\gamma_1 = \gamma R$. The regression parameter $\xi$ can be uniquely identified from the joint distribution of $(X,Y)$. 

As discussed in Section 2.1, $\beta = \xi - \gamma\delta$. Let $\delta_1 = \arg \min_{\delta} ||\xi -\gamma_1  \delta||_0$ and $\beta_1 =  \xi - \gamma_1 \delta_1$ denote solutions to the optimization problem (5). Therefore, we have $||\beta_1||_0 \leq ||\beta||_0 \leq (p-q)/2$. Consequently, the difference between $\beta_0$ and $\beta_1$ must have at least $(p+q)/2 + (p+q)/2 - p = q$ zero entries. According to (2), $\xi = \beta+\gamma \delta = \beta_1 + \gamma_1\delta_1$, and $\beta - \beta_1 = \gamma_1\delta_1 - \gamma\delta = \gamma(R\delta_1 - \delta)$. Since any submatrix of $\gamma$ consisting of $q$ rows has full rank under assumption $A3$, and that $\gamma(R\delta_1 - \delta)$ has at least $q$ zero entries, we can establish that $R\delta_1 - \delta = 0$, which indicates $\beta_1 = \beta$. Therefore, we conclude that $\beta$ is identifiable through the optimization problem (5).



\section{Misspecification of $q$ in low dimensional simulations}
\label{supp:q}

\begin{figure}[htbp!]
\centering
\begin{subfigure}[b]{.45\textwidth}
    \centering
    \includegraphics[width=\textwidth]{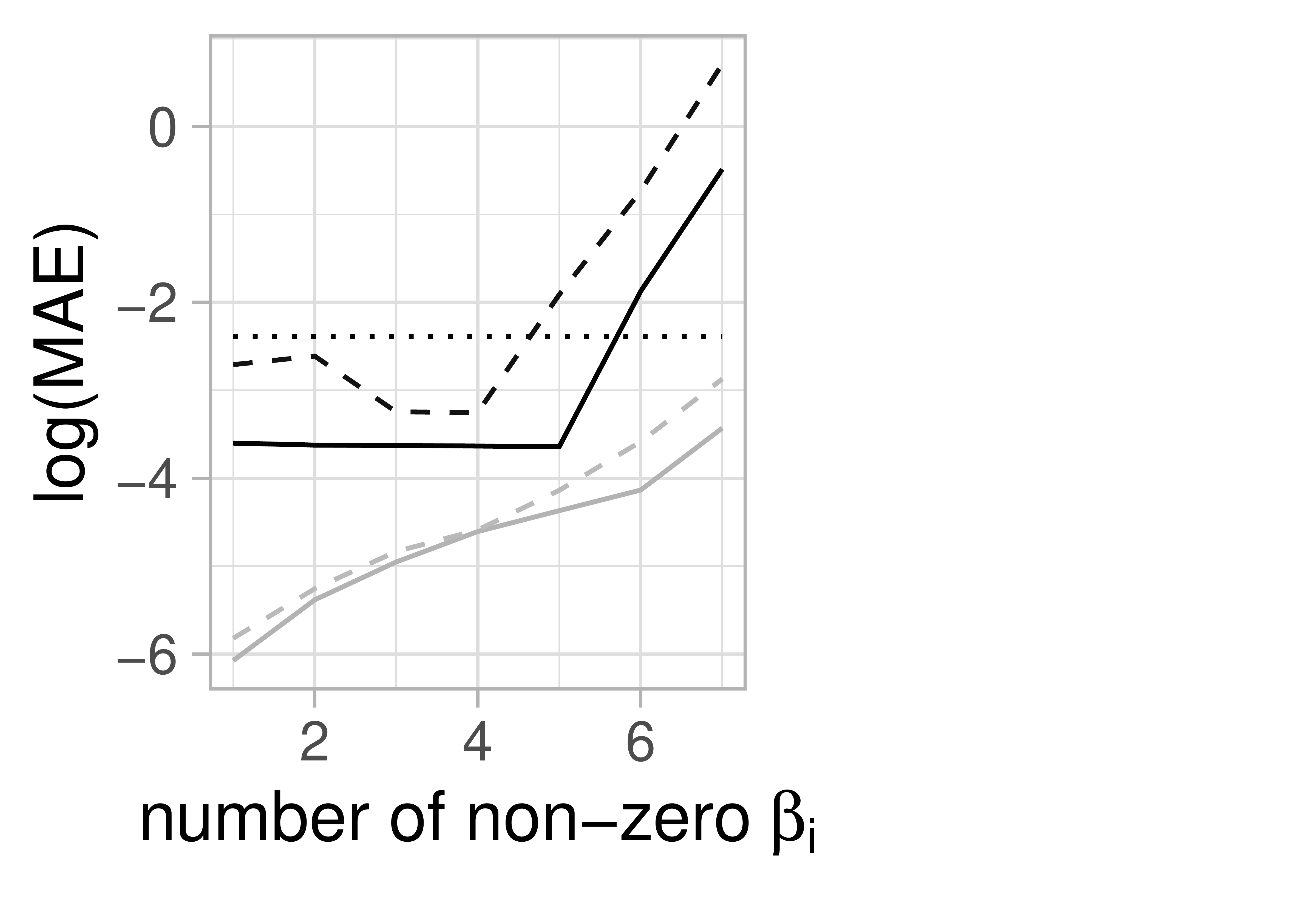}
\end{subfigure}\hspace{-3em}%
\begin{subfigure}[b]{.45\textwidth}
    \centering
    \includegraphics[width=\textwidth]{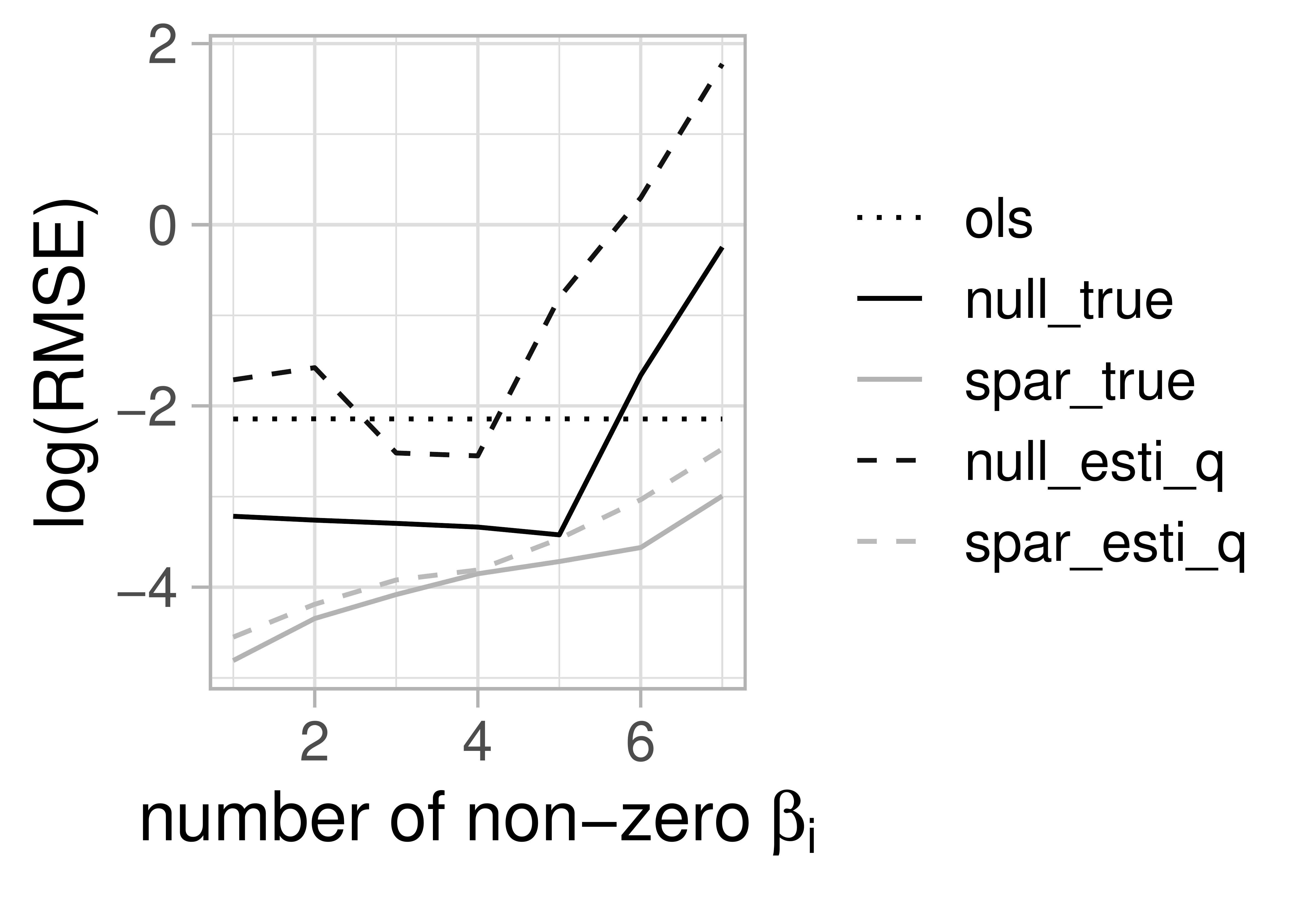}
\end{subfigure}
\caption{The MAE and RMSE of the Spar, Null and OLS under correct specification of number of unmeasured confounders (denoted as true), and estimated number of latent confounders (denoted as esti\_q).}
\label{fig:errbysparse_estiq}
\end{figure}

Figure \ref{fig:errbysparse_estiq} illustrates the estimation errors for both the Spar and Null methods under correct specifications of the number of unmeasured confounders $q$ and cases where $q$ is estimated through the hypothesis testing method in \cite{nfactOnatski}. Utilizing estimated numbers of latent factors results in higher estimation errors for both Spar and Null methods compared to cases where the true number of unmeasured confounders is utilized. However, the influence of estimating $q$ on the Spar method is negligible. Furthermore, the estimation errors of the Spar method under the estimated $q$ remain smaller than those of the Null method under the correct specification of $q$. The superior performance of our proposed Spar method is evident regardless of whether the number of unmeasured confounders is known a priori or estimated.

\begin{figure}[htbp!]
\centering
\begin{subfigure}[b]{.45\textwidth}
    \centering
    \includegraphics[width=\textwidth]{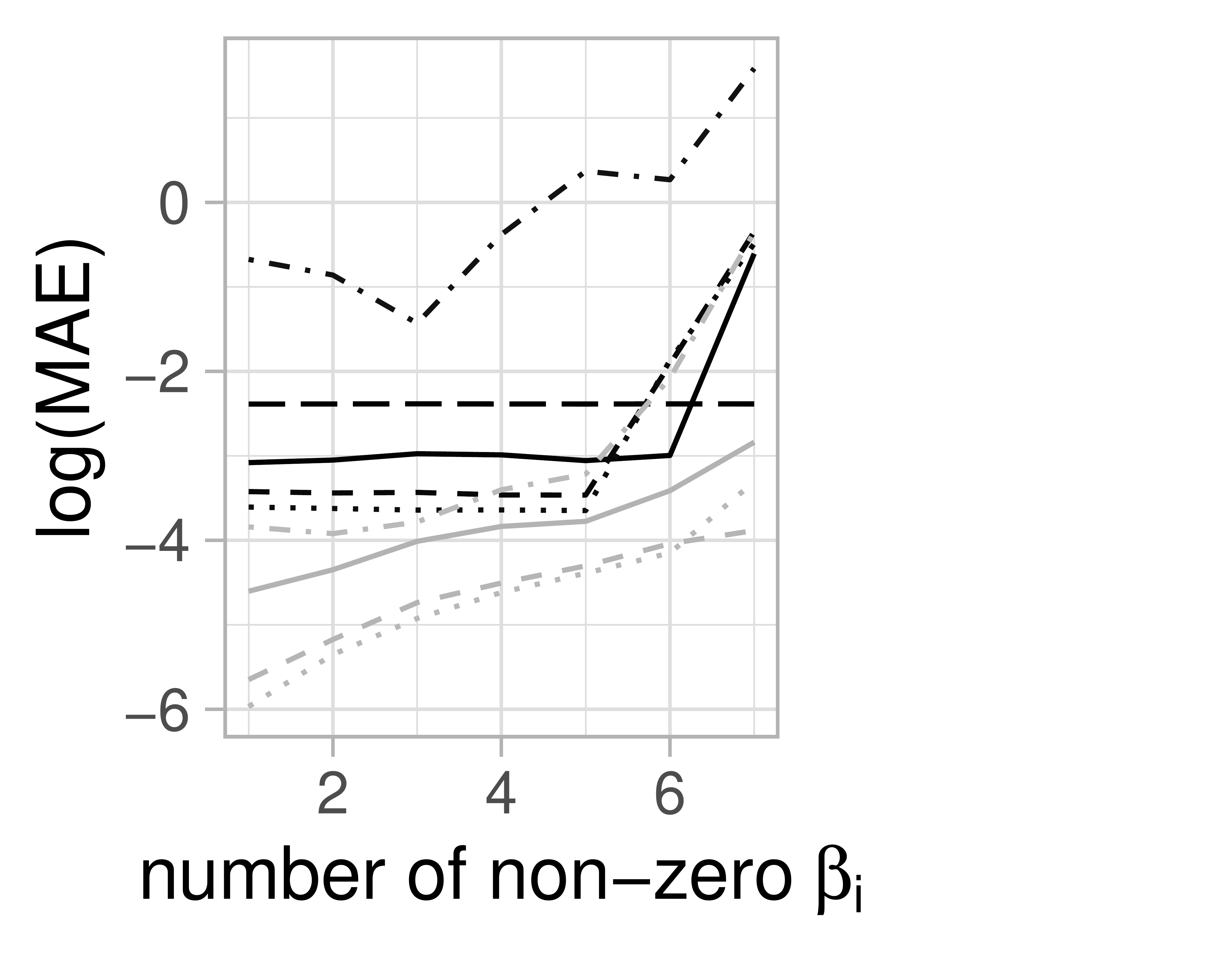}
\end{subfigure}\hspace{-1em}%
\begin{subfigure}[b]{.45\textwidth}
    \centering
    \includegraphics[width=\textwidth]{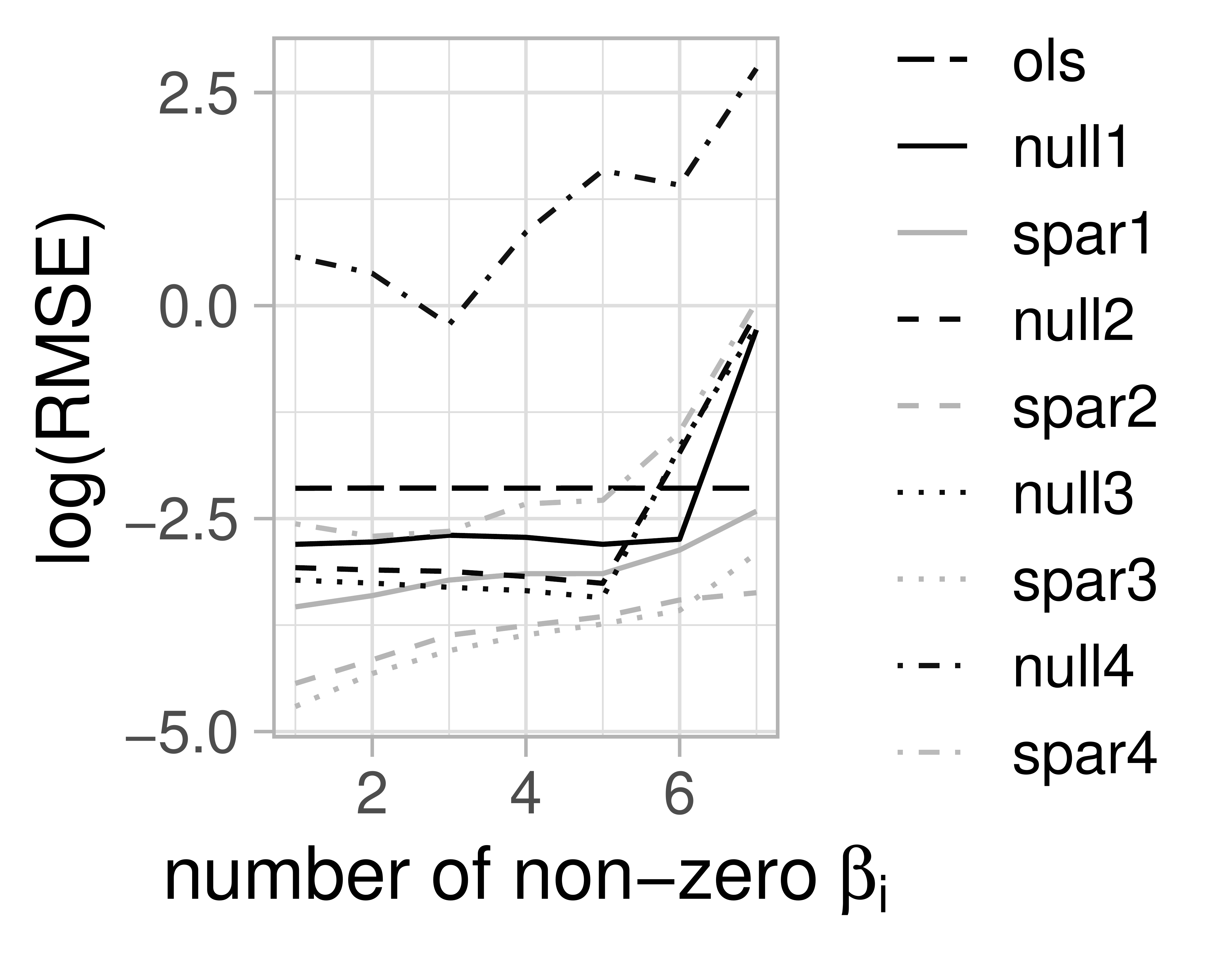}
\end{subfigure}
\caption{The MAE and RMSE of the Spar, Null and OLS methods under different sepecifications of number of unmeasured confounders $q$. The $x$-axis represents various sparsity levels of $\beta$. The numbers in the legend indicate the specified number of latent confounders. $q=3$ is the true number of confounders.}
\label{fig:errbysparsenfact}
\end{figure}

Figure \ref{fig:errbysparsenfact} shows the estimation errors of the Spar and Null methods across different numbers of latent factors $q$. Notably, the estimation errors are minimized when $q=3$. Both the Spar and Null methods are sensitive to overspecified $q$ values (e.g. when $q=4$). Conversely, when the specified $q$ is smaller but proximate to the true value (e.g. $q=2$), the estimation errors for both methods show modest increases compared to cases where $q$ is correctly specified. In summary, in the context of low-dimensional scenarios, both the Spar and Null methods are sensitive to the specification of the number of unmeasured confounders. Practically, it is advisable to prioritize underestimating rather than overestimating the number of latent confounders. Furthermore, it is recommended to assess the robustness of estimation by varying the specified number of latent confounders in practical applications.

\section{Simulation models of $S$ and $\gamma$}
\label{supp:gwas}

In Section 4, we generate $S$ and $\Gamma$ under the following three scenarios.

\begin{enumerate}
    \item Balding-Nichols Model (BN): $d = 3$. Each column $i$ in $\Gamma$ consists of i.i.d. draws from the Balding-Nichols model: $\Gamma_{ki} \overset{\text{i.i.d.}}{\sim} \text{BN}(p_i, F_i)$, where $k \in {1,2,3}$. The pairs $(p_i, F_i)$ are computed by randomly selecting a SNP from the HapMap dataset, calculating its observed allele frequency, and estimating its $F_{ST}$ value using the Weir \& Cockerham estimator \citep{BNsim}. The rows of $S$ are sampled i.i.d. from $\text{Multinomial}(60/210, 60/210, 90/210)$, reflecting the subpopulation proportions in the HapMap dataset.
    \item Pritchard-Stephens-Donnelly (PSD): $\Gamma$ is simulated similarly to the BN model. The rows of $S$ are sampled i.i.d. from a Dirichlet distribution with parameters $(0.5,0.5,0.5)$.
    \item Spatial: $d = 3$. Each column $i$ in $\Gamma$ is simulated with i.i.d. draws: $\Gamma_{ki} \overset{\text{i.i.d.}}{\sim} 0.9 \times \text{Uniform}(0,0.5)$ for $k = 1,2$, and $\Gamma_{3,i} = 0.05$. The first two columns of $S$ are i.i.d. samples from a Beta distribution with parameters $(\tau, \tau)$, where $\tau = 0.1$, while the third column of $S$ is set to be 1.
\end{enumerate}

\section{Violation of sparsity assumption in synthetic GWAS data}
\label{supp:disturbance}

\begin{figure}[htbp!]
\centering
\includegraphics[width=1\textwidth]{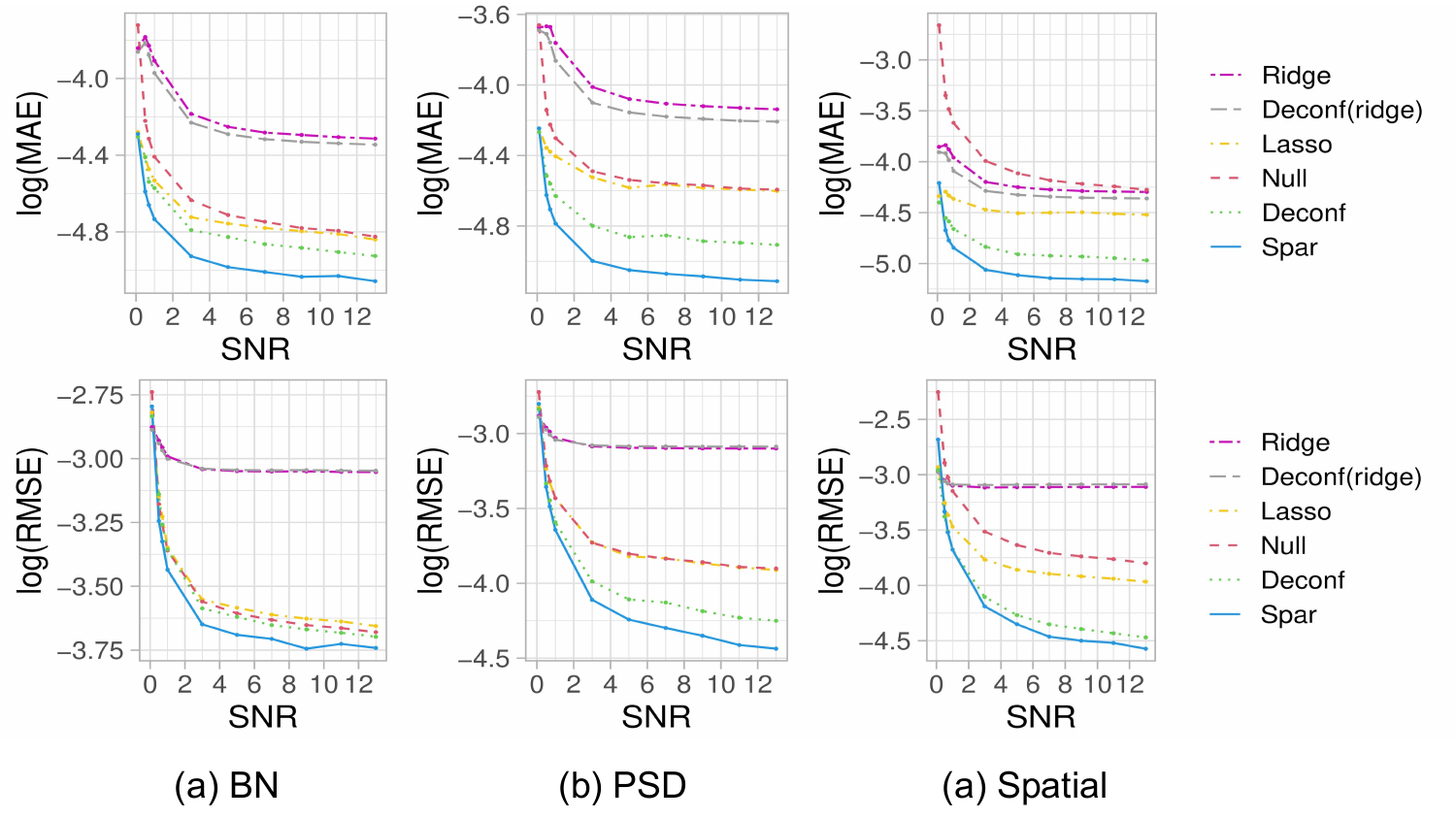}
\caption{MAE and RMSE of the method Spar, Null, Deconf, Deconf(ridge), Lasso, and Ridge under different signal-to-noise rates, in the case when minor disturbances are introduced for $\beta$.}
\label{fig:noise}
\end{figure}

The assumption $A1$ requires that the causal effects of the majority of the treatments should be zero. In our previous synthetic GWAS date generations, we assumed that the causal effects of the last 99\% genes were precisely zero, which may not fully capture the complexity of real-world scenarios. In reality, many treatments may have weak but non-zero causal effects. To account for such occasions, we assign a uniform distribution to the last 99\% causal effects $\beta \sim Uniform$(-0.05,0.05) while keeping all other simulation settings unchanged. The resulting MAE and RMSE are presented in Figure \ref{fig:noise}. The error trends with increasing SNR closely resemble the patterns observed when the sparsity assumption holds. The Spar method consistently exhibits the lowest MAE and RMSE among all the methods evaluated. Notice that the MAE for both the Lasso and Spar methods increases when minor disturbances are introduced for $\beta$, as compared to the case when $\beta$ is sparse, which is expected since both methods yield sparse estimators. Our findings demonstrate the robustness of the Spar method in addressing real-world scenarios where the sparsity assumption may not be strictly met.

\section{Simulation results with measured confounders $W$}
\label{supp:measuredw}
\begin{figure}[htbp!]
\centering
\begin{subfigure}[b]{.45\textwidth}
    \centering
    \includegraphics[width=\textwidth]{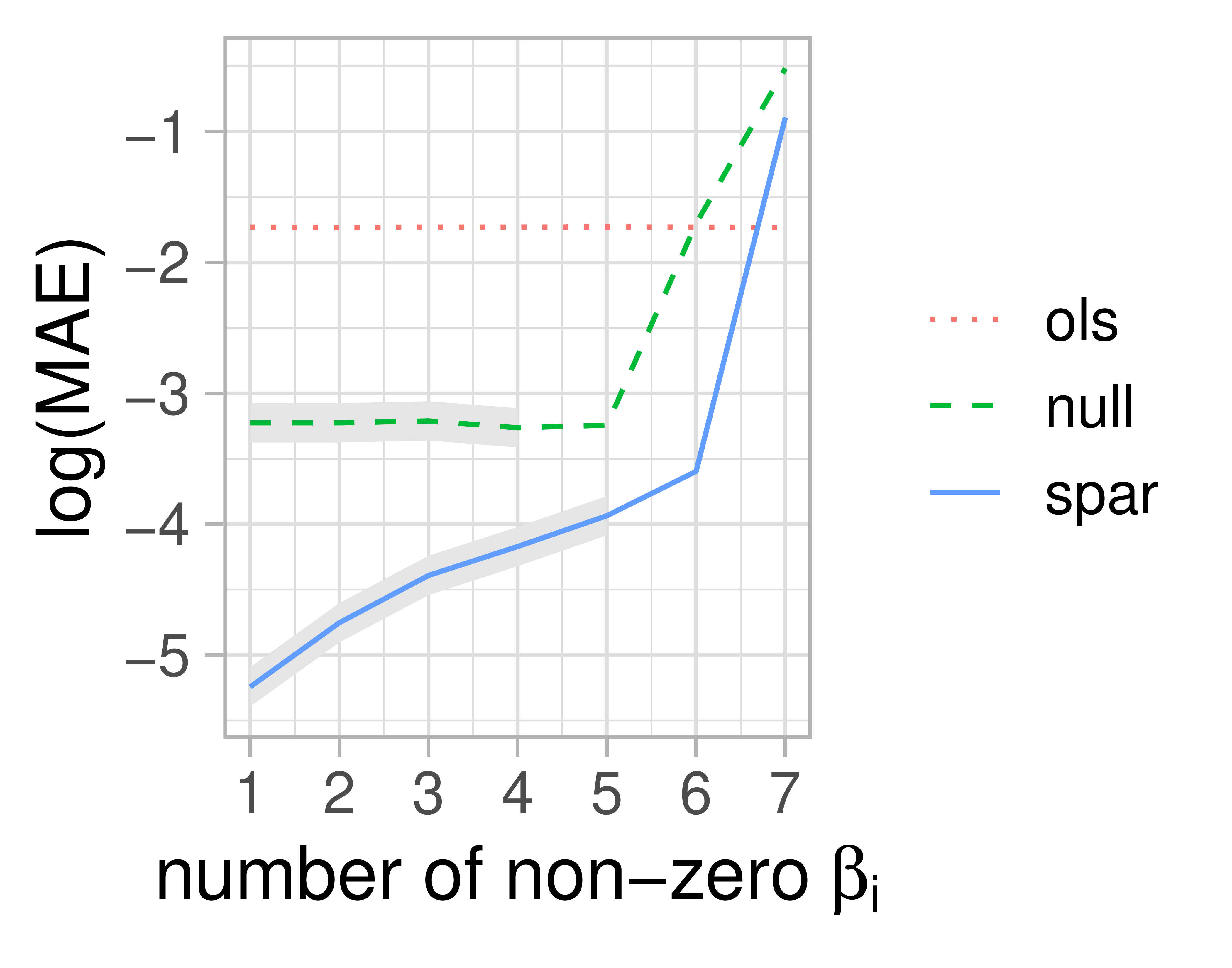}
\end{subfigure}\hspace{-1em}%
\begin{subfigure}[b]{.45\textwidth}
    \centering
    \includegraphics[width=\textwidth]{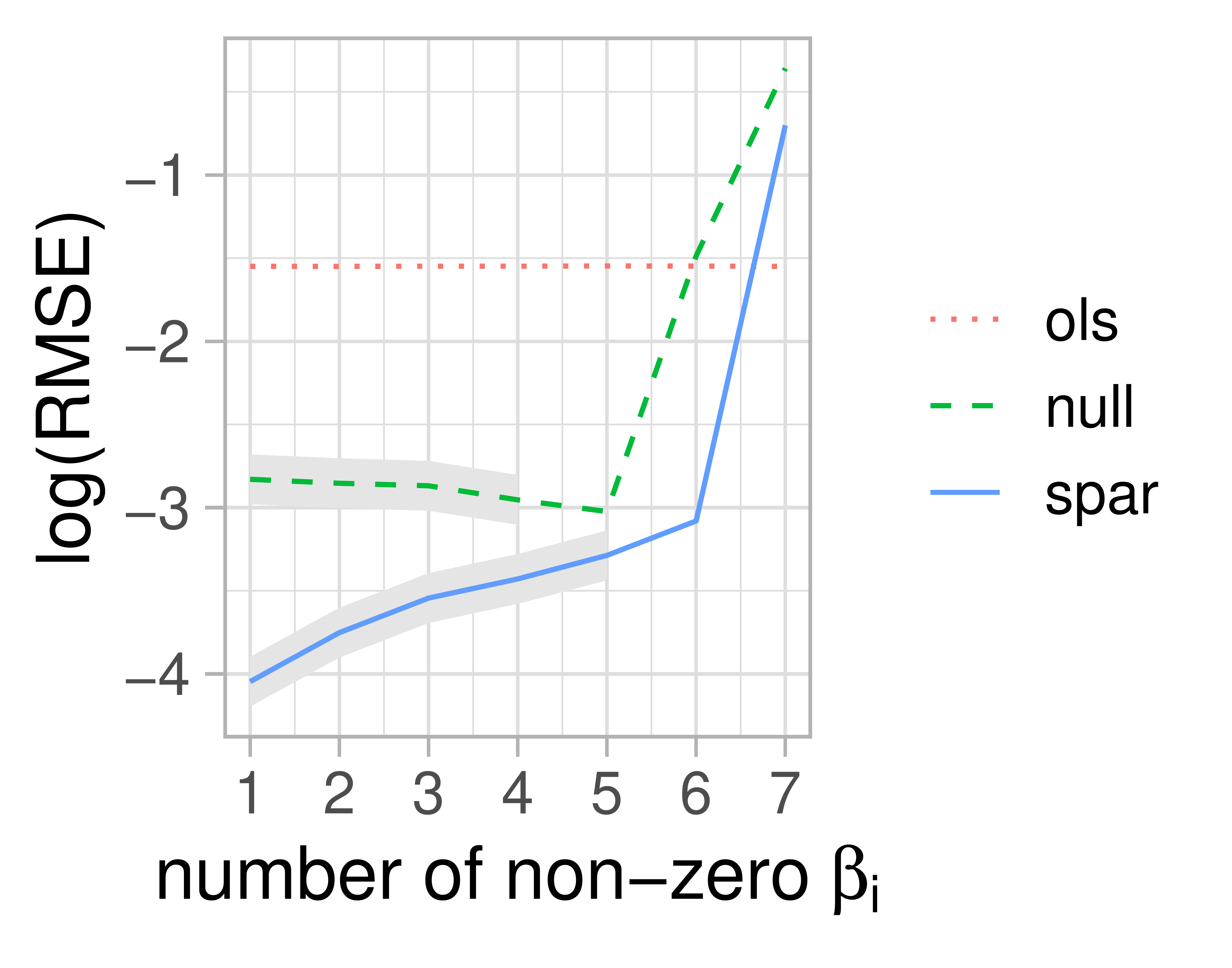}
\end{subfigure}
\caption{The MAE and RMSE for the Spar, Null, and OLS methods with a 3-dimensional measured confounder.}
\label{fig:errbysparsewithw}
\end{figure}

In this section, we reran the simulations in both low and high dimensional settings as described in Section 3, incorporating the measured confounder \(W\). For the low-dimensional setting, we added a 3-dimensional variable \(W\) simulated from \(N(0, I_3)\). We set \(\lambda = (1, 1, 1)^T\) and randomly generated the entries of \(\eta\) from \(Uniform(-1, 1)\), keeping all other simulation settings identical to those in Section 3.1. For estimation, we first regressed both \(X\) and \(Y\) on the measured confounder \(W\), and then used the residuals from the linear regression as new outcomes and treatments for the estimation procedure. Figure \ref{fig:errbysparsewithw} presents the MAE and RMSE of both the Spar and Null methods across different sparsity levels of the causal effects \(\beta\). The results indicate that the performance of our method remains consistent, whether the measured variable is included or not.

\begin{figure}[htbp!]
\centering
\begin{subfigure}[b]{.45\textwidth}
    \centering
    \includegraphics[width=\textwidth]{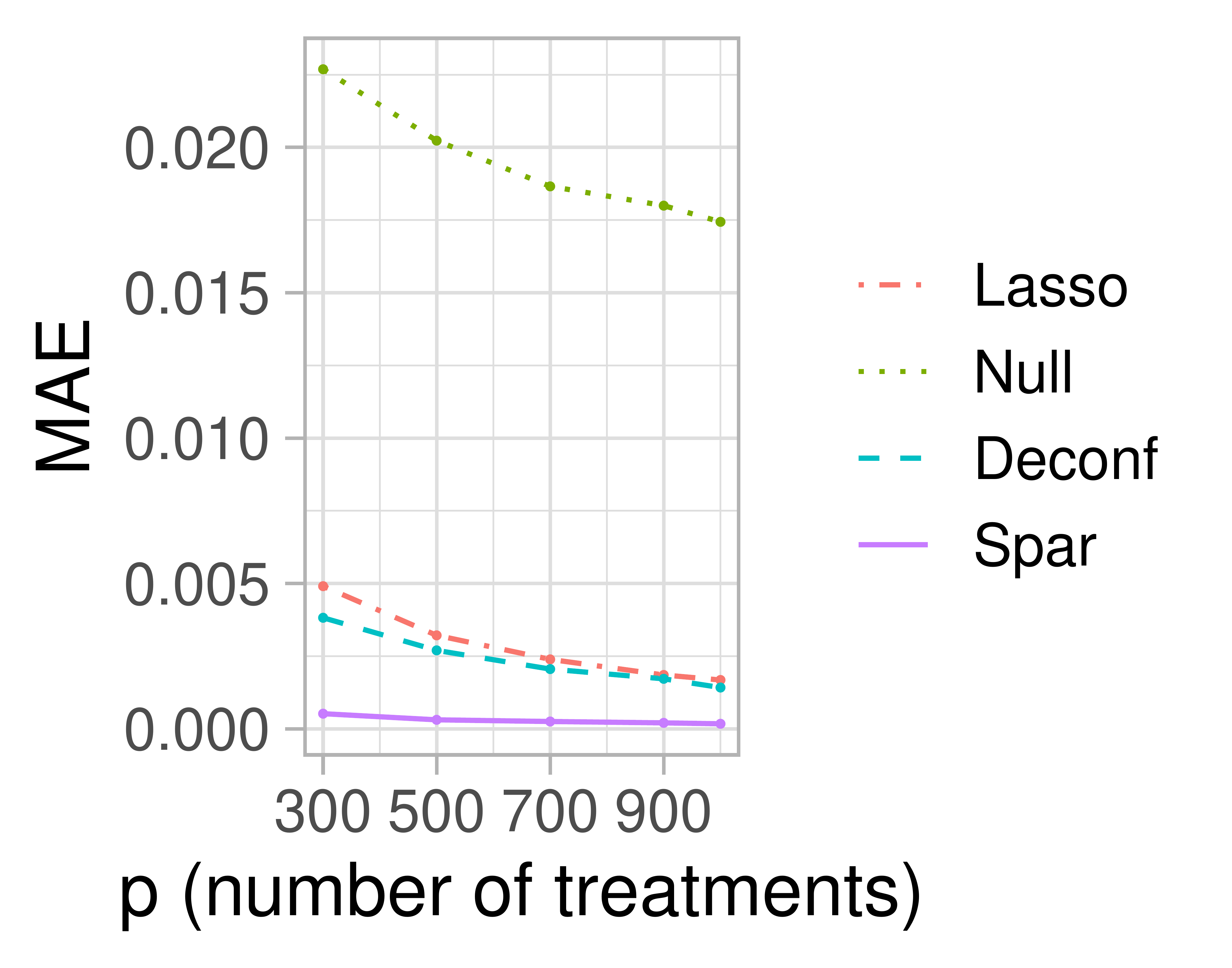}
\end{subfigure}%
\begin{subfigure}[b]{.45\textwidth}
    \centering
    \includegraphics[width=\textwidth]{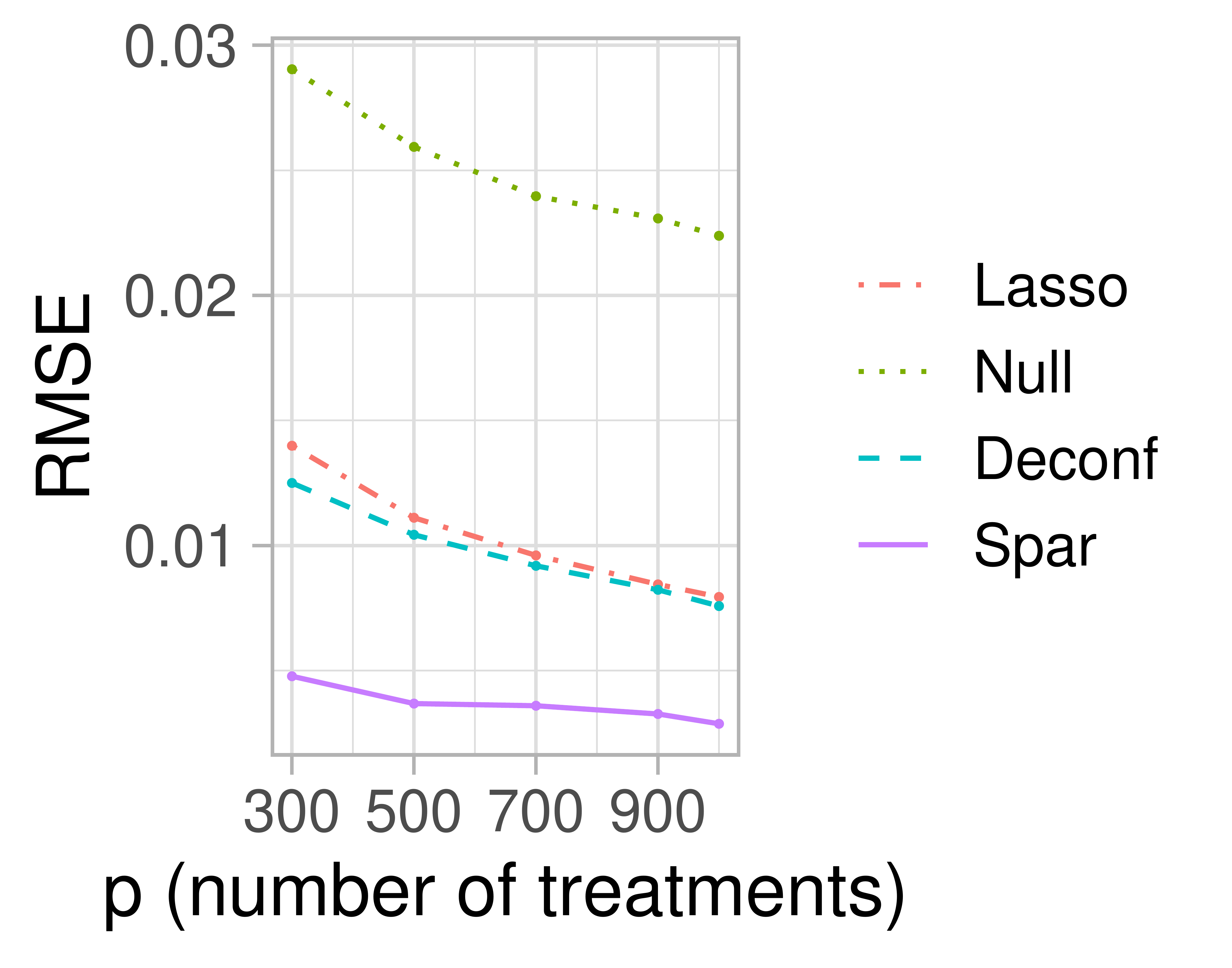}
\end{subfigure}
\caption{The MAE and RMSE of Spar, Null, Deconf and OLS methods under different number of treatments when $n=300$. The dimension of measured variable $r=3$.}
\label{fig:errorhighd3}
\end{figure}

\begin{figure}[htbp!]
\centering
\begin{subfigure}[b]{.45\textwidth}
    \centering
    \includegraphics[width=\textwidth]{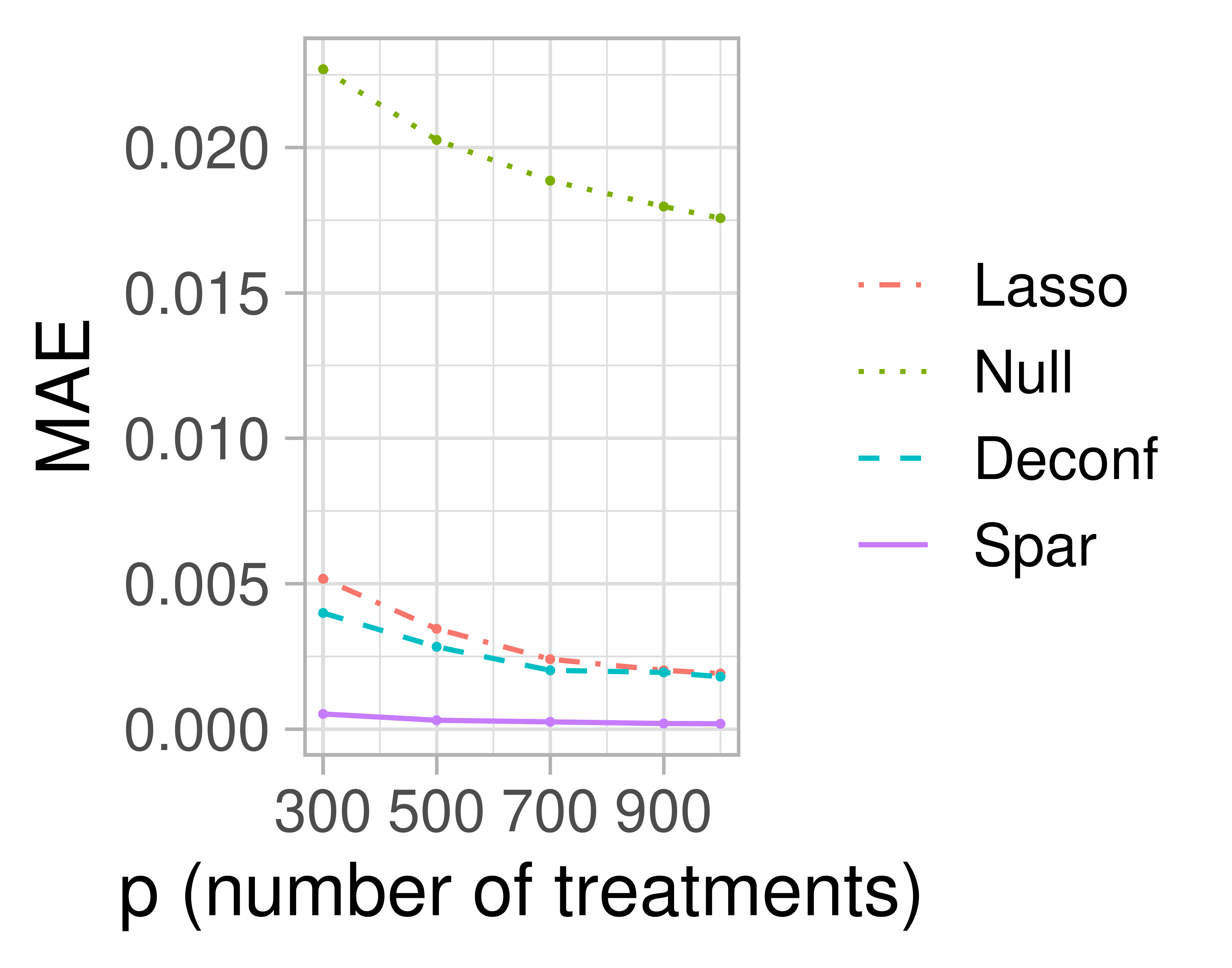}
\end{subfigure}%
\begin{subfigure}[b]{.45\textwidth}
    \centering
    \includegraphics[width=\textwidth]{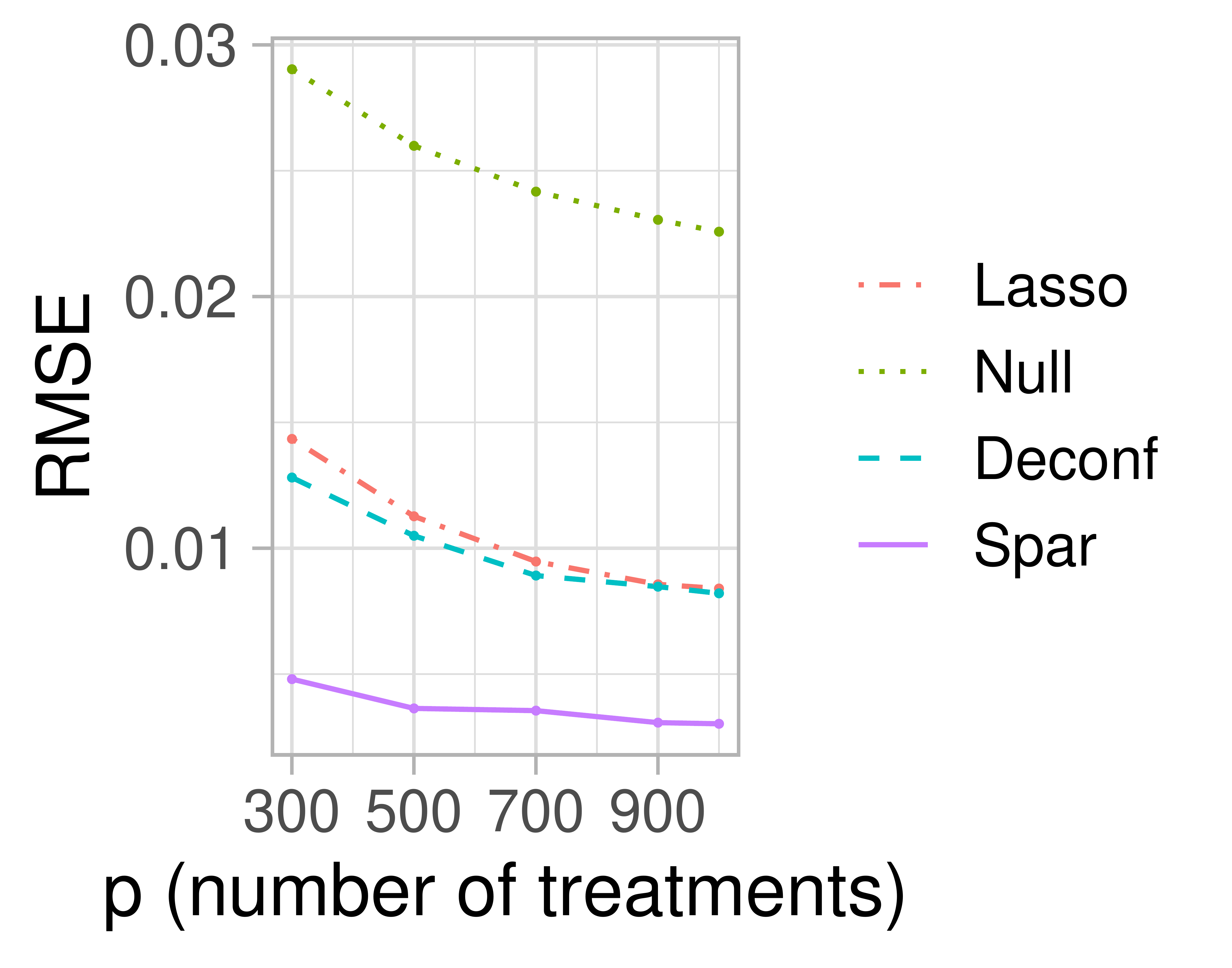}
\end{subfigure}
\caption{The MAE and RMSE of Spar, Null, Deconf and OLS methods under different number of treatments when $n=300$. The dimension of measured variable $r=10$.}
\label{fig:errorhighd10}
\end{figure}

\begin{figure}[htbp!]
\centering
\begin{subfigure}[b]{.45\textwidth}
    \centering
    \includegraphics[width=\textwidth]{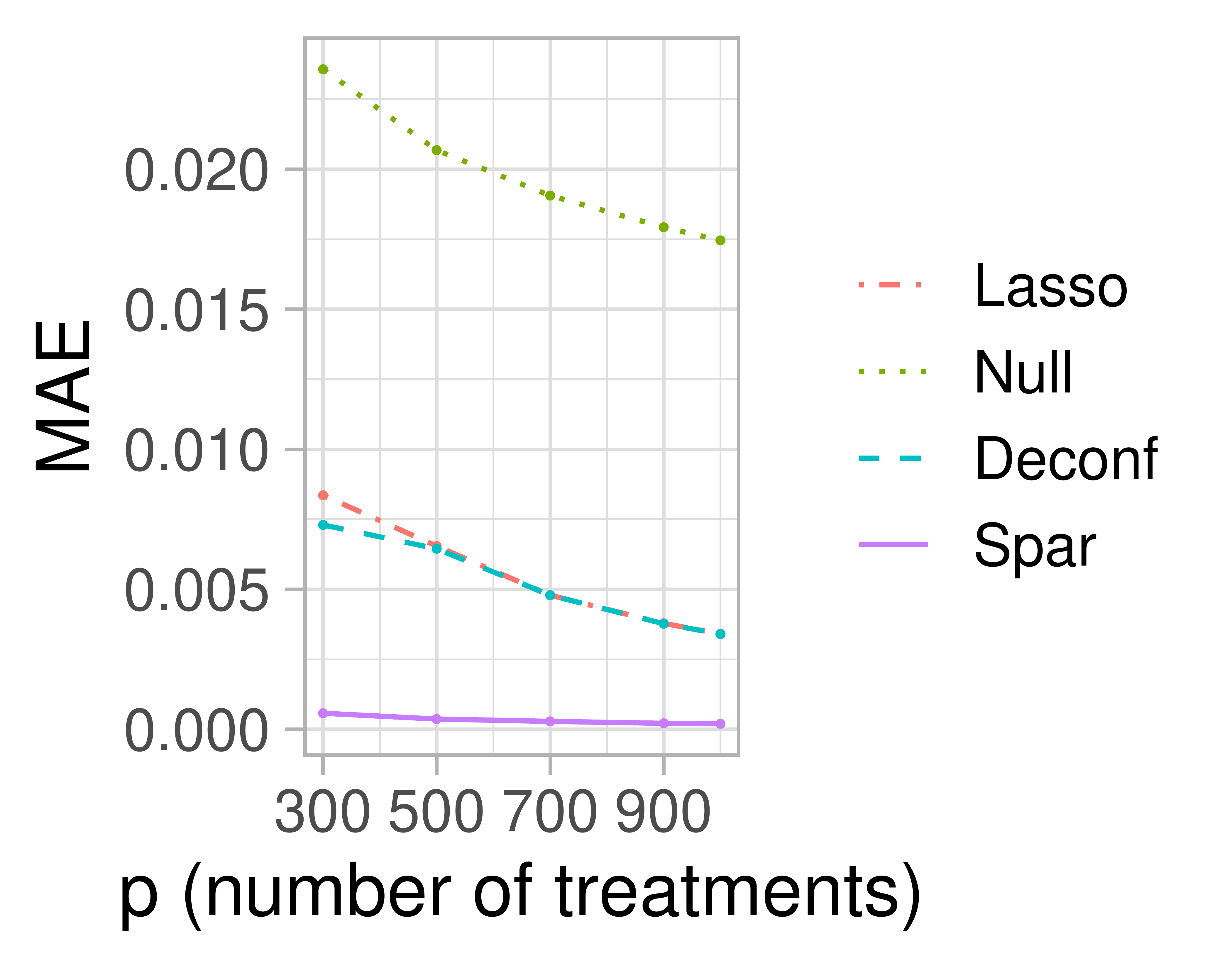}
\end{subfigure}%
\begin{subfigure}[b]{.45\textwidth}
    \centering
    \includegraphics[width=\textwidth]{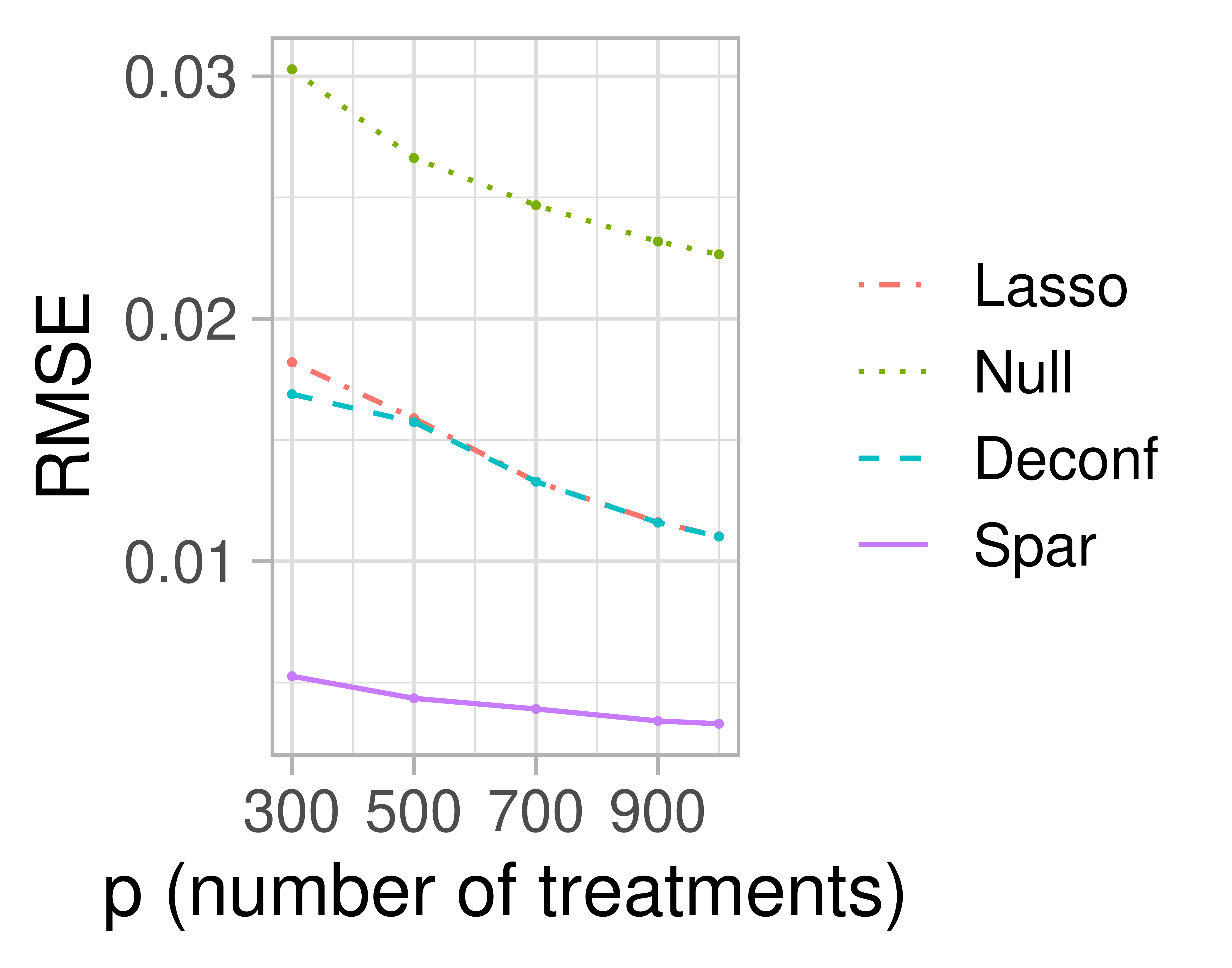}
\end{subfigure}
\caption{The MAE and RMSE of Spar, Null, Deconf and OLS methods under different number of treatments when $n=300$. The dimension of measured variable $r=50$.}
\label{fig:errorhighd50}
\end{figure}

In the high-dimensional setting, we included measured confounders with dimensions 3, 10, and 50, respectively. We simulated the measured variable \(W\) from \(N(0, I_r)\), where \(r = 3, 10, 50\). Each element of \(\eta \in \mathcal{R}^{p \times r}\) and \(\lambda \in \mathcal{R}^r\) was generated i.i.d. from \(Uniform(-1, 1)\). All other simulation settings remained consistent with those in Section 3.2. Figures \ref{fig:errorhighd3}, \ref{fig:errorhighd10}, and \ref{fig:errorhighd50} display the MAE and RMSE for measured confounder dimensions of 3, 10, and 50, respectively. The results demonstrate that our method performs consistently, regardless of the dimension of the measured variables.

\newpage

\bibliographystyle{chicago}
